\documentclass[fleqn,usenatbib]{mnras}

\usepackage[T1]{fontenc}
\usepackage{ae,aecompl}

\usepackage{graphicx}	
\usepackage{amsmath}	
\usepackage{amssymb}	
\usepackage{physics}
\usepackage{placeins}
\usepackage{cleveref} 
\usepackage[swedish, english]{babel}
\usepackage{newtxtext,newtxmath}
\usepackage{comment}
\newcommand{\ang}{\,\textup{\AA}}
\newcommand{\Msol}{\,M_{\odot}}
\graphicspath{ {figures/} }

\title[Conditions for detecting Pop III galaxies]{Conditions for detecting lensed Population III galaxies in blind surveys with the {\it James Webb Space Telescope}, the {\it Roman Space Telescope} and {\it Euclid}}

\author[Anton Vikaeus et al.]{
Anton Vikaeus,$^{1}$\thanks{E-mail: anton.vikaeus@physics.uu.se}
Erik Zackrisson,$^{1}$
Daniel Schaerer,$^{2,3}$ 
Eli Visbal,$^{4}$ \newauthor
Emma Fransson,$^{1}$ 
Sangeeta Malhotra,$^{5}$ 
James Rhoads,$^{5}$ 
Martin Sahlén.$^{6,7}$ 
\\
$^{1}$Observational Astrophysics, Department of Physics and Astronomy, Uppsala University, Uppsala, Sweden \\
$^{2}$Geneva Observatory, University of Geneva, Chemin Pegasi 51, CH-1290 Versoix, Switzerland \\
$^{3}$CNRS, IRAP, 14 Avenue E. Belin, 31400 Toulouse, France \\
$^{4}$University of Toledo, Department of Physics and Astronomy and Ritter Astrophysical Research Center, 2801 Bancroft, Toledo, OH 43606, USA \\
$^{5}$Astrophysics Science Division, NASA Goddard Space Flight Center, 8800 Greenbelt Road, Greenbelt, Maryland, 20771, USA \\
$^{6}$Swedish Collegium for Advanced Study, Thunbergsvägen 2, SE-752 38 Uppsala, Sweden \\
$^{7}$Theoretical Astrophysics, Department of Physics and Astronomy, Uppsala University, Uppsala, Sweden \\
}
\date{Accepted XXX. Received YYY; in original form ZZZ}

\pubyear{2021}

\begin{document}
\label{firstpage}
\pagerange{\pageref{firstpage}--\pageref{lastpage}}
\maketitle

\begin{abstract}
Dark matter halos that reach the HI-cooling mass without prior star formation or external metal pollution represent potential sites for the formation of small -- extremely faint -- Population III galaxies at high redshifts. Gravitational lensing may in rare cases boost their fluxes to detectable levels, but to find even a small number of such objects in randomly selected regions of the sky requires very large areas to be surveyed. Because of this, a small, wide-field telescope can in principle offer better detection prospects than a large telescope with a smaller field of view. Here, we derive the minimum comoving number density required to allow gravitational lensing to lift such objects at redshift $z=5-16$ above the detection thresholds of blind surveys carried out with the \textit{James Webb space telescope} (\textit{JWST}), the \textit{Roman space telescope} (\textit{RST}) and \textit{Euclid}. We find that the prospects for photometric detections of Pop III galaxies is promising, and that they are better for \textit{RST} than for \textit{JWST} and \textit{Euclid}. However, the Pop III galaxies favoured by current simulations have number densities too low to allow spectroscopic detections based on the strength of the HeII1640 emission line in any of the considered surveys unless very high star formation efficiencies ($\epsilon \gtrsim 0.1$) are envoked. We argue that targeting individual cluster lenses instead of the wide field surveys considered in this paper results in better spectroscopic detection prospects, while for photometric detection, the wide field surveys perform considerably better.
\end{abstract}




\section{Introduction}
The very first stars -- the so-called Population III (hereafter Pop III) -- formed from pristine gas containing only the elements produced during big bang nucleosynthesis (H, He and trace amounts of Li), and are generically expected to be very massive, with typical masses in the  $\sim 10$--1000 $M_\odot$ range \citep[e.g.][]{Hosokawa16,Hirano18}. At the end of their lifetimes, these objects are expected to either enrich the surrounding interstellar medium by means of supernovae or mass lost on the asymptotic giant branch \citep[e.g.][]{Heger02,Rydberg2020}, or directly collapse into black holes \citep{Latif2020,2020Whalen}. In the case of very low mass \citep{2018Magg} or low formation redshift \citep{2020LiuBromm}, Pop III stars could in principle also survive to the present era. 

Detecting Pop III stars remain one of the milestone observations needed to piece together the puzzle of early structure formation, but individual Pop III stars are expected to be too faint, by a large margin, to be directly observed with present or nascent facilities, unless subject to extreme gravitational lensing \citep{Rydberg13,2018Windhorst}. The exception would be the rare case of a supermassive ($M\sim 10^5 M_\odot$) Population III star \citep{Haemmerle18}. These gargantuan stars, considered already in the 1960s as direct sources of the, by then, newly discovered quasi-stellar objects \citep[although argued as dynamically unstable by e.g.,][]{chandrasekhar1964} now also offer a venue of direct detection of Pop III stars, their supernovae or their resulting direct collapse black holes with only mild magnifications or even without the help of gravitational lensing \citep{Surace2018,2019Surace,2020Whalen}.

In this article we pursue the alternative avenue of detecting the integrated light from many Pop III stars formed simultaneously within the same Pop III galaxy\footnote{sometimes alternatively referred to as Pop III star clusters} \citep[e.g.][]{2008Johnson,2009Johnson,2010Stiavelli,2017Visbal,2017Yajima,2018Inayoshi,2019Johnson,2019Kulkarni}, i.e., the summed light from several hundreds to thousands of Pop III stars formed within a time window of just a few million years inside a single, chemically pristine dark matter halo. The detection of such objects could potentially grant us insights into the stellar initial mass function (IMF) of Pop III stars, the number densities of atomic cooling dark matter halos that host Pop III galaxies, the Pop III star formation efficiency ($\epsilon$), large scale metal-mixing etc.. 

The observational signatures typically associated with Pop III galaxies are related to the fact that a top-heavy IMF produces a very hard ionizing flux because of the high effective temperatures of Pop III stars. This outgoing flux will ionize and excite the surrounding interstellar medium, producing strong hydrogen and helium emission lines such as Ly$\alpha$, He II $\lambda$1640, H$\alpha$ and H$\beta$ without corresponding metal lines -- indicative of a completely pristine top-heavy stellar population \citep{2002Schearer,2003Schearer,2011Inoue,2011Zackrisson}.

Summing up the light from several hundreds to thousands of Pop III stars surely improves our prospects for detecting some of the very first stellar structures in the Universe. Unfortunately this, in itself, will generally not suffice for the Pop III galaxies seen in current simulations, unless we at the same time may take advantage of large magnifications due to gravitational lensing from foreground objects \citep[]{2012Zackrisson,2015Zackrisson}.

In this paper we will investigate the prospects for upcoming telescopes to detect Pop III galaxies with the help of gravitational lensing in blind surveys targeting redshifts in the range $z=5$--16, i.e. from a cosmic epoch prior to the predicted peak of the Pop III star formation rate density at $z\sim 10$ \citep[e.g.][]{2020LiuBromm} to just after the end of cosmic reionization.  

Assessing the detectability of gravitationally lensed Pop III galaxies requires us to determine the probabilities of achieving a certain gravitational magnification, where the minimum required magnification depends on the intrinsic brightness of the high-redshift Pop III galaxies that we are searching for. With this in hand we can calculate the number densities\footnote{Throughout this paper we will refer to the number of Pop III galaxies per comoving cubic megaparsec simply as ''number density'' unless stated otherwise.} (number of Pop III galaxies per comoving Mpc$^{3}$) required to allow the detection of some minimum number of highly magnified Pop III galaxies in the survey area probed by future space telescope missions. These number densities can then be compared to those predicted in contemporary simulations of Pop III galaxies. 

Throughout this paper, we focus exclusively on ``pure'' Pop III galaxies, i.e. objects without any metal-enriched stars. Simulations indicate that Pop III stars could potentially also form inside pockets of pristine gas within an otherwise metal-enriched galaxy \citep[e.g.][]{Sarmento18,Sarmento19}, thereby giving rise to a spectrum containing a mixture of both Pop III and Pop II/I spectral signatures, but the detectability of such ``hybrid'' Pop III galaxies is beyond the scope of this paper.

Throughout the paper, we use a flat $\Lambda$CDM cosmological model with the following parameters throughout the paper; $H_0=67.3\mathrm{\, km \, s^{-1}Mpc^{-1}}$, $\Omega_{\Lambda}=0.685$, $\mathrm{\Omega_{m}=0.315}$ and $\mathrm{\Omega_b=0.0487}$ \citep{2014Planck}.

The paper is organized as follows; In Sect.~\ref{method} we describe the model framework in which we perform our calculations, and the assumptions involved. In Sect.~\ref{results} we present our findings and in Sect.~\ref{discussion} discuss their implications. Our conclusions are summarized in Sect.~\ref{conclusions}.

\section{Method \& Models}
\label{method}
The vast majority of Pop III galaxies will, due to their low intrinsic luminosities and large distances from us, remain hidden below the detection thresholds of even the largest telescopes currently planned \citep{2020Schauer}. For every Pop III galaxy, there is however a small but non-zero probability for strong gravitational lensing from matter overdensities (galaxies or galaxy clusters) along the line of sight, which could lift the object into the detectable regime. This probability is a function of the magnification and
the redshift of the object, although the latter dependence is relatively weak throughout the high-redshift regime considered  here (see fit in eq.~\ref{eqPfit}). Hence, a given survey volume must contain a minimum number of Pop III galaxies in order to ensure a reasonable probability that at least a few are sufficiently magnified to end up above the detection threshold of that survey. This may be converted into a lower limit on the comoving number density of Pop III galaxies (with a given intrinsic luminosity) at the survey redshift that Pop III galaxies must meet to ensure reasonable detection prospects. This limit may then be compared to the properties (comoving number density and luminosity) of Pop III galaxies predicted by numerical simulations to assess whether gravitational lensing has any chance of rendering them observable in upcoming surveys of the high-redshift Universe. In this section, we present the computational machinery for deriving this limit. The focus here is on large-area surveys (many square degrees) that would inevitably cover both galaxy clusters and void regions, or alternatively small-area surveys (a few square arcminutes) that cover random fields. Small-area surveys that focus exclusively on strong-lensing galaxy cluster fields are outside the scope of this paper \citep[but see][for a study of this type on the detectability of Pop III galaxies]{2012Zackrisson}.  

\subsection{Spectral synthesis modelling}
\label{SED}
The data for the Pop III galaxy emission line strengths and luminosities used for our predictions are taken from the model by \cite{2010Raiter}. In this paper we consider a fiducial top-heavy lognormal stellar initial mass function (IMF) of characteristic mass $m_c = 60\, M_\odot$ and width $\sigma = 1$, in the mass range $1 - 500 \, M_\odot$ that is forming stars at a constant rate (as in the \cite{2006Tumlinson} lognormal case E IMF). Pop III galaxies are likely to retain their tell-tale spectral signatures only during a short time after the initial starburst, before supernova feedback quenches further star formation and the surrounding nebula is polluted by metals, we therefore consider a maximum lifetime $t_\mathrm{Pop III} = 10$ Myr as our fiducial Pop III galaxy ``lifetime'' and is therefore used as the star forming age ($\tau$) at which we extract the spectrum.

\subsection{Spectral signatures}
In this paper, we primarily consider two different criteria for the detection of Pop III galaxies in blind surveys -- detection of their rest-frame UV ($1500 \, \ang$) continuum flux using photometry, or detection of their He II $\lambda$1640 line emission using spectroscopy.
 
 The $1500 \, \ang$ continuum detection criterion is the less demanding of the two, and simply means that the Pop III galaxy has been rendered sufficiently bright by gravitational lensing to end up above the detection threshold of a photometric survey (typically a multi-band survey from which dropouts can be used to assess redshift) and would hence be expected to appear in its source catalog. However, meeting this criterion does not necessarily imply that the Pop III galaxy would be identifiable among the other types of far more numerous sources detectable in the same survey. This $1500 \, \ang$ flux limit nonetheless serves as an indicator of whether the galaxies could in principle be targeted for followed-up spectroscopy (either with \textit{JWST} or a ground based large telescope). Identifying a Pop III galaxy based on photometry alone may not be impossible although this would require special conditions to be met. This includes colour signatures due to strong Balmer line emission yet missing metal lines (requires rest-frame optical data, hence \textit{JWST} photometry; \citealt{2011Inoue,2011Zackrisson}), colour signatures due to very strong Ly$\alpha$ emission (requires high Ly$\alpha$ emission through the intergalactic medium, i.e. that the Pop III galaxy is sitting in an ionized bubble or is at $z\lesssim 6$; \citealt{2003Pello,2011Zackrissonb}), colour signatures due to the Lyman bump (requires a redshift $z<6$ to avoid significant absorption by the intergalactic medium; \citealt{2010Inoue}) and colour signatures due to a very blue UV continuum slope (requires extreme leakage of ionizing photons to reveal the otherwise subdominant stellar continuum; \citealt{2010Raiter,2011Zackrisson}). We stress that for some of these signatures, it seems unrealistic to assume that the required physical conditions would be met by more than a small fraction of the Pop III galaxies in a given survey volume, which could make the compounded probability of having a galaxy with just the right colours signature for identification {\it and} a fortunate sightline for extreme magnification, very low.

The He II $\lambda$1640 emission line originates from gas regions surrounding sources of highly energetic radiation, due to the high ionization potential of helium. Owing to the expected top-heavy IMF and high temperatures of Pop III stars, one suspects that a strong He II $\lambda$1640 emission line (i.e. a high equivalent width; bright line with respect to the surrounding continuum) should be indicative of a Pop III galaxy in cases when no other metal lines can be observed \citep[e.g.][]{2002Schearer,2003Schearer,2010Raiter}. 
For our fiducial Pop III galaxy IMF and star-forming age ($\tau\sim 10$ Myr, see section ~\ref{SED}), the predicted He II $\lambda$1640 emission line luminosity is given by $\sim 6\times 10^{41}$ erg s$^{-1}$, normalized to SFR $= 1 \ \Msol$/yr. At $z\sim10$ such a Pop III galaxy has an emission line flux of $\sim 4.5\times 10^{-22}$ erg s$^{-1}$ cm$^{-2}$ -- for a total stellar mass $M_\star \sim 10^{4} \Msol$. The equivalent width is found to be about $24.5 \,\ang$, which is considerably stronger than that seen in other metal-poor galaxies -- the next to lowest metallicity used in \cite{2010Raiter} (i.e., $Z=5\times 10^{-6} \,Z_\odot$) dramatically reduces the equivalent width to a mere $2.8 \, \ang$. Observations of metal-poor galaxies have typically returned equivalent widths in the range $1-3 \, \ang$ \citep{2019Nanayakkara,2020Saxena,2020Feltre}. In \cite{2020Saxena}, at $z=2.5-5$, they find equivalent widths reaching at most $ 8 \, \ang$ when removing galaxies harbouring a suspected active galactic nuclei (AGN) from the observed sample (one of the seven suspected AGN included in the data set had an equivalent width of $21.4 \,\ang$ while the other six showed equivalents widths $<7.5\,\ang$). On the other hand, in \cite{2019Nanayakkara}, at $z=2-4$, one finds a few objects with equivalent widths reaching the levels of the Pop III stellar populations in our fiducial models. That is, one may still find metal-enriched galaxies that are strong He II $\lambda$1640 emitters with equivalent widths reaching that expected of a pristine Pop III galaxy with a top-heavy IMF. However, taking into account other sources of helium ionizing radiation, such as AGN activity, and removing them from the sample, observations typically return equivalent widths that are significantly lower than for a Pop III galaxy. Hence, one expects this emission line to be a very powerful diagnostic of potential Pop III galaxy candidates. 

One should note, however, that there are still considerable uncertainties attached to the expected $1500 \, \ang$ continuum fluxes, He II $\lambda$1640 emission-line fluxes and He II $\lambda$1640 equivalent widths for Pop III galaxies, even in the case where the Pop III IMF is assumed to be known. As discussed by \citet{2010Raiter}, the continuum fluxes at $1500 \, \ang$ and $1640 \, \ang$ (the latter of which affects the He II $\lambda$1640 equivalent width)  are not dominated by direct star light, but by nebular 2-photon continuum, which is sensitive to the assumed gas density (with high density leading to a reduction in flux), whereas the He II $\lambda$1640 line flux is sensitive to the assumed ionization parameter of the gas (with low ionization parameter leading to a reduction in He II line flux due to absorption of He$^+$-ionizing photons by H atoms, unlike what is expected from simplified recombination models). The \citet{2010Raiter} models are, moreover, based on non-rotating Pop III stars, and as shown in simulations by \citet{2011Stacy,2013Stacy}, one typically expects Pop III stars to be rapid rotators, unless significantly affected by magnetic breaking \citep{2018Hirano}. \citet{2012Yoon} argues that rotation may significantly boost the He$^+$-ionizing fluxes of Pop III stars by factors of several, which could lead to a corresponding increase in He II $\lambda$1640 line emission. At low total stellar masses ($\lesssim 10^4 \ M_\odot$), the He II $\lambda$1640 line flux may furthermore be significantly affected by IMF sampling effects \citep{2016MasRibas}, which increases the variance of this quantity throughout the Pop III galaxy population. As argued by \citet{2020Vikaeus}, even slightly metal-enriched galaxies with star formation rates (SFR) $\lesssim 1 \, M_\odot$ yr$^{-1}$ may show elevated He II $ \lambda 1640$Å equivalent widths due to IMF sampling issues, to the point that a low-metallicity galaxy without Pop III stars may come across as a Pop III hybrid galaxy (although not likely a pure Pop III galaxy of the type we discuss here). In summary,  both the $1500 \, \ang$ continuum flux and the He II $\lambda$1640 emission line are likely uncertain by a factor of a few, even for a fixed functional form of the Pop III stellar IMF. While our fiducial models are based on \citet{2010Raiter} predictions, in section~\ref{results} we also discuss how of factor of $\approx 3$ departures from these predictions would affect our conclusions.

\subsection{Magnification distribution}
The magnification data were taken from \cite{2015Zackrisson}, where magnification ($\mu$) probability functions ($P(>\mu)$) were calculated using a ray-tracing algorithm on a sequence of lens planes populated with both dark matter from N-body simulations, and galaxies based on a semi-analytical model. \cite{2020Robertson} presented a comparison of high-magnification ($\mu>10$) probabilities based on similar models from different groups, and at the redshift limit of their comparison ($z\approx 5$), our results agree well (differing only on the order of a few per cent) with those of \cite{2008Hilbert} and \cite{2008Tinker}. At lower redshifts ($z\gtrsim 1$) our model agrees well with \cite{2020Robertson} and \cite{2008Hilbert} -- at most reaching relative variations $\sim 20$ per cent, while \cite{2008Tinker} consistently have probabilities $\sim 50$ per cent higher than our model at these low redshifts. In this paper we consider Pop III galaxy redshifts in the range $z = 5-16$, hence the low redshift discrepancies are not directly relevant for the stringency of our results.

We find that, for magnifications $\mu > 10$ and redshifts in the range $z=3-16$, the functional form of our magnification data can be approximated by a power-law in $\mu$ with a weak polynomial dependence on redshift (in the range considered):
\begin{equation} \label{eqPfit}
    P(>\mu) \approx \left(-4.2\times 10^{-5}z^2 + 1.3\times 10^{-3}z + 1.5\times 10^{-4} \right)\mu^{-2}.
\end{equation}
As argued in \cite{2015Zackrisson,1982Peacock} we cannot, however, apply the magnification probability function to arbitrary magnifications ($\mu$) for extended objects such as star clusters or galaxies. As we consider larger magnifications, the size of the region in the lens plane responsible for the magnification maps to an increasingly small area in the source plane. For $\mu > 1000$ and a source plane redshift of $z=10$ we begin to encounter differential magnification on scales $\lesssim 10$ pc, for which reliable predictions would require detailed models for the spatial distribution of the stars in the galaxy and the surface brightness profile of the ionized gas. This in particular limits the usage of the \textit{RST} and \textit{Euclid} surveys to a minimum galaxy mass ($M_{\star.\mathrm{min}}$), while for \textit{JWST} this will only be an issue when doing spectroscopy or photometry on very low mass galaxies ($\lesssim 1000 \ \Msol$) at high redshift. The minimum galaxy mass applicable for a given telescope can be estimated by using eq.~\ref{eqmuPhoto} or eq.~\ref{eqmuSpec} while setting $\mu_\mathrm{min}\geq 1000$ and solving for the mass. For the  wide (deep) \textit{RST} and \textit{Euclid} surveys, using $z=10$ and star-forming age $\tau \approx 10$ Myr, this puts us at minimum total stellar masses; $3\times10^4\, M_\odot$ $(4.8\times10^3\, M_\odot)$ and $4.8\times10^5\, M_\odot$ $(7.5\times10^4\, M_\odot)$ respectively for photometry, and at $2.2\times10^6\, M_\odot$ $(2.2\times10^5\, M_\odot)$ and $6.7\times10^6\, M_\odot$ $(1.1\times10^6\, M_\odot)$ respectively for spectroscopy. Similarly, for \textit{JWST} one finds; $1.1\times10^3\, M_\odot$ and $2.9\times10^3\, M_\odot$ for photometry and spectroscopy respectively (based on the deep NIRCam and very deep NIRSpec observing modes presented in table~\ref{surveyparams}). We indicate this minimum allowed galaxy mass in figure~\ref{fig_Tele} for all the different telescopes and survey modes by a star shaped marker that is color coded to match the respective telescope. In particular, the survey modes with low depth are limited from below in the galaxy masses that they can detect, even with strong gravitational lensing. Throughout the paper, a redshift $z=10$ and a star forming age $\tau\approx10$ Myr are chosen as a fiducial values due to the star formation rate density of Pop III stars peaking around $z=10$ \citep[e.g.,][]{2020LiuBromm} and the spectra of the galaxies simulated in this paper reaching stable values after a couple of million years which will be retained throughout the defined lifetime of 10 Myr before forming distinctive spectral features from metal enrichment. The predictions in this paper which relate to total stellar mass therefore refers to Pop III galaxy that has formed their stars during a 10 Myr window -- which can then be compared to simulations with predicted Pop III total stellar masses forming over a similar time window. If we detect a Pop III galaxy we should note it may very well have any age within the time window between its formation and the upper limit of 10 Myr. Since the luminosity of the galaxy is determined by the SFR (for a given IMF) we may find Pop III galaxies with total stellar masses that are lower than the mass at which we set the detection limit, and they would still be observable. In other words, the SFR determines the detectability while the allowed time window sets the maximum total stellar mass that we expect to observe.

\subsection{Population III galaxy simulations}
As the first stars form through gravitational collapse out of pristine gas, devoid of any metals that enhance cooling, we must consider environments in which hydrogen can cool effectively via, either atomic line transitions or molecular hydrogen cooling. Without such cooling mechanisms, the collapse of pristine gas will halt and resist star formation for very extended times \citep{2013Bromm}. We therefore require the formation of dark matter halos of sufficient mass in which the virial temperature of the pristine gas reaches $\mathrm{T_{vir}\approx 10^4 K}$ whereupon HI atomic cooling may commence and thus ensure a promising formation site for a Pop III galaxy. As star formation ramps up in the Universe, a background of Lyman-Werner (LW) radiation is built up. This background radiation is able to suppress star formation in lower-mass dark matter halos -- at redshift $z=10$, the LW background is sufficiently high that essentially all star formation occurs in halos with a mass corresponding to the atomic cooling limit, or in more massive halos \citep{2020Visbal}. This provides us with a typical dark matter halo mass, which in turn -- depending on the average cosmic density of baryonic matter with respect to dark matter and the star formation efficiency $\epsilon$ of pristine gas -- gives us an estimate of the typical stellar mass of a Pop III galaxy across different redshifts.

\begin{figure*} 
  \centering
    \includegraphics[width=\columnwidth]{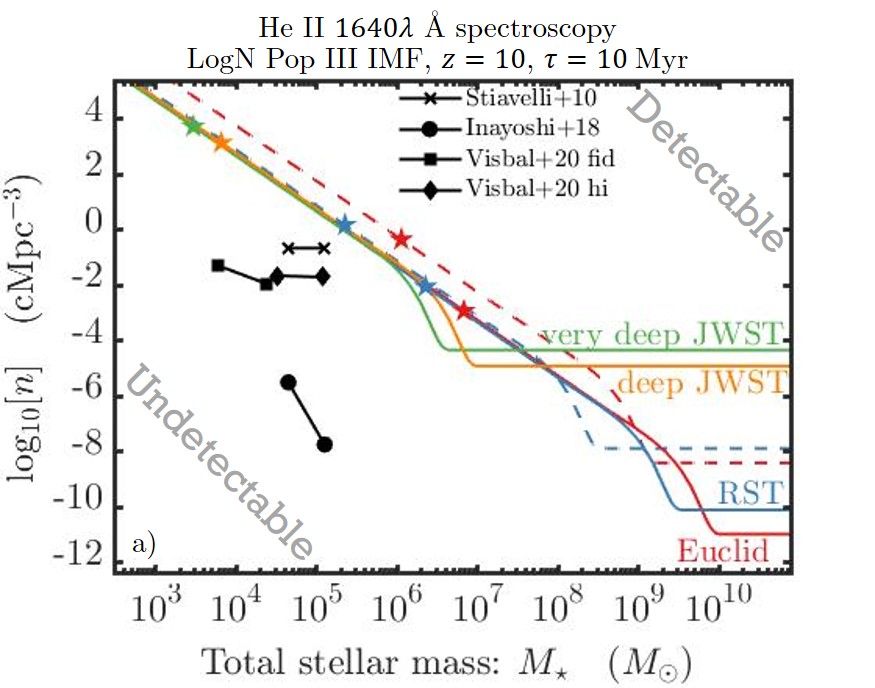}
    \includegraphics[width=\columnwidth]{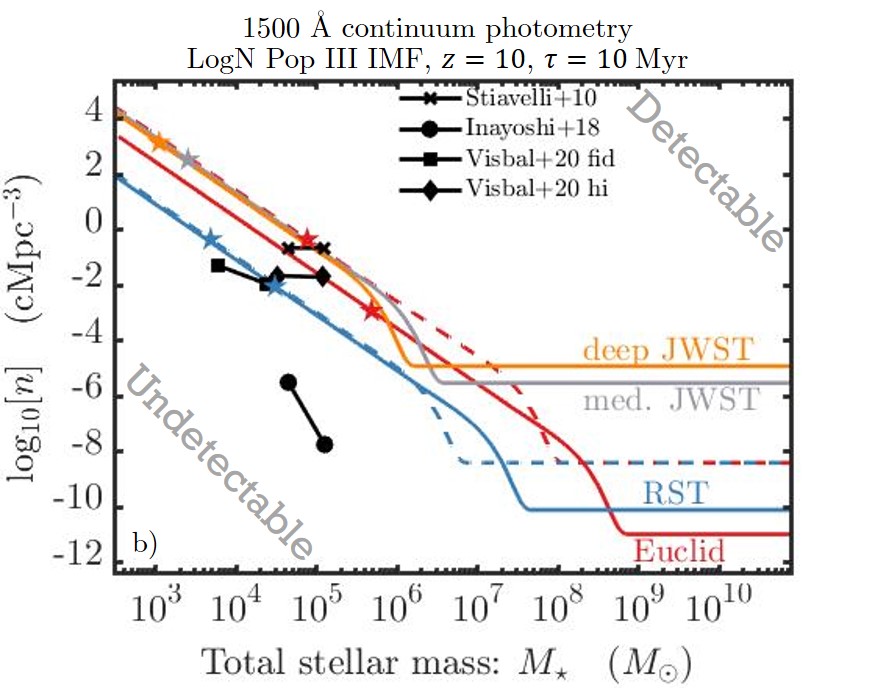}
    \caption{ a): The minimum number density $(n_\mathrm{_{min}})$ of Pop III galaxies required for spectroscopic detection of the He II $\lambda$1640 $\ang$ emission line as a function of the total stellar mass of the Pop III galaxy, at redshift $z=10$. b): The minimum number density $(n_\mathrm{_{min}})$ required for photometric detection of the continuum at 1500 $\ang$ as a function of the total stellar mass of the Pop III galaxy, at redshift $z=10$. The stellar population synthesis model used \protect\citep{2010Raiter} in the above figures is based on a top-heavy lognormal IMF ($m_c=60 \,M_\odot$, $\sigma = 1$, with mass range $1 - 500 \, M_\odot$) with a constant star formation rate where the spectrum is extracted when the galaxy has reached a star forming age of $\tau = 10$ Myr. The lines correspond to different telescopes with survey parameters as specified in table \ref{surveyparams}, the dashed blue and red lines are for deep \textit{RST} and deep \textit{Euclid} surveys respectively. The black marked points are the results of simulations predicting the number density of atomic cooling halos (with an allowed temperature range $1\times10^4-2\times10^4$ K) which may host Pop III galaxies with masses calculated using $\epsilon = 0.01$ for \protect\cite{2010Stiavelli,2018Inayoshi} and $\epsilon = \{0.001,0.005\}$ for the \protect\cite{2020Visbal} fiducial (fid) and high (hi) efficiency model respectively. The lower masses in the \protect\cite{2020Visbal} fiducial and high efficiency model corresponds to the atomic cooling limit while the slightly higher masses indicates halos that have star formation suppressed to higher masses due to photo-heating as a result of cosmic reionization. We have indicated the minimum mass ($M_\mathrm{\star,min}$) allowed in order not to breach the $\mu_\mathrm{max}=1000$ limit for each telescope by the star symbols, color coded to their respective telescopes. The figures can be crudely separated by the colored lines into two regions, the detectable and the undetectable region. Any Pop III galaxy comoving number density located in the upper-right region, above the required minimum comoving number density set by the telescopes, would have a good chance of ending up above the detection limits of future surveys. On the contrary, Pop III galaxy comoving number densities located underneath the required minimum, would have a small likelihood of being detected. At $z=10$, the figure reveals that all the telescopes have almost identical prospects for spectroscopic detection of the He II $\lambda$1640 emission line in galaxies with mass $M_\star \lesssim 10^6 \, M_\odot$ -- the only exception being the deep \textit{Euclid} survey (red dashed line). \textit{JWST} has an advantage in very deep exposures when it comes to the most massive Pop III galaxies of total stellar mass $M_{\star} \gtrsim 10^6 M_\odot$. When it comes to photometric detection of the UV-continuum at 1500 Å, the best telescope is a wide or deep \textit{RST} survey for all Pop III galaxy masses in the relevant mass range. Trailing the wide and deep \textit{RST} surveys is the wide \textit{Euclid} survey, we also find that there is a notable area around $M_\star \approx 10^6\, M_\odot$ where \textit{JWST} can be competitive in deep and medium-deep surveys. Based on the minimum required number density of the different surveys as compared to the simulations, we conclude that for the given star formation efficiency, spectroscopic detection is unlikely. When it comes to photometry, the \protect\cite{2010Stiavelli} model very likely provide detections with \textit{RST} and likely also the wide \textit{Euclid} survey. The \protect\cite{2020Visbal} high efficiency model is most likely to provide detections with \textit{RST} and in rare cases also \textit{Euclid}.}\label{fig_Tele}
  \end{figure*}

To make proper use of our estimates for the required minimum number density of Pop III galaxies, we make comparisons with a number of independent theoretical simulations that attempt to predict the number density of Pop III galaxies at various redshifts. We use the predictions made by \cite{2010Stiavelli,2018Inayoshi} and \cite{2020Visbal} in order to determine whether the required number densities are indeed reproduced in contemporary simulations. \cite{2010Stiavelli} calculates the formation rate of pristine atomic cooling halos per cMpc$^3$ in the temperature range $10^4 \leq T_\mathrm{vir} \leq 2\times 10^4$ K subject to a homogeneous Lyman-Werner background as the mechanism delaying star formation, and adopt the assumption of inhomogeneous metal mixing to estimate the efficiency with which halos are enriched and thus lose their pristine Pop III observational signatures. The range in virial temperature implies a range in the total stellar mass of Pop III galaxies, given an assumed star formation efficiency, which can be seen in figures \ref{fig_Tele} and \ref{fig_IMFvar}. \cite{2020Visbal} similarly provides us with number densities and total Pop III stellar masses but for different assumed star formation efficiencies, taking into account an inhomogeneous H$_2$ dissociating Lyman-Werner background, HI ionizing background from inhomogeneous reionization and inhomogeneous metal enrichment feedback processes. This provides a more dynamical and realistic formation scenario for Pop III galaxies, yielding number densities that are lower than \citet{2010Stiavelli} by about an order of magnitude. The discrepancy between the models can be explained by the more refined physical treatment in \citet{2020Visbal}  which has the effect of suppressing the number density. Variations in the parameters underlying the mechanisms for delayed star formation was shown in \citet{2020Visbal} to only affect the predicted abundance of Pop III stars by a factor of a few, therefore underscoring the robustness of the models. Furthermore, the star formation efficiency is a free parameter in these models which, in the \citet{2020Visbal} fiducial case, is assumed to not reach significantly above $\epsilon = 0.001$. We also include a high efficiency model from \citet{2020Visbal} which allow for the formation of slightly more massive Pop III galaxies ($\epsilon = 0.005$). The true value for the Pop III star formation efficiency is debated and has not been constrained well yet due to the absence of large high-z ($z\gtrsim 10$) datasets. In addition to our fiducial star formation efficiency, we have therefore allowed for several alternatives in figure~\ref{fig_nminVSz} to visualise the need to achieve high star formation efficiencies. We have also included \cite{2018Inayoshi} which simulates the number density of pristine atomic cooling halos forming through frequent mergers in areas of high baryonic streaming motion as the process delaying star formation. These kind of formation processes produce fewer suitable atomic cooling halos in which pristine Pop III galaxies can form but might allow for higher star formation efficiencies similar to what is found in \citet{2020Regan} which simulates rapidly forming halos. As a bottom line, we regard the \citet{2010Stiavelli} and \citet{2020Visbal} simulations as viable estimates of number density of pristine Pop III galaxies while \citet{2018Inayoshi} provide lower bounds through more rare formation channels.

Using these predictions we estimate the comoving number density of these atomic cooling halos for the corresponding total stellar mass forming in such halos via eq.~\ref{eqMvir} \& \ref{eqMstellar} for the \citet{2010Stiavelli} and \citet{2018Inayoshi} simulations. \citet{2020Visbal} predicts similar stellar masses that differ slightly due to modulation by the local LW flux. The virial mass of the atomic cooling halo for the given cosmology is taken to be, as given by \cite{1998Bryan&Norman,2001Barkana&Loeb}:
\begin{equation} \label{eqMvir}
    M_\mathrm{vir} \approx 32.7 \times T_\mathrm{vir}^{3/2}\left(\frac{1+z}{10}\right)^{-3/2}
\end{equation}
for a neutral primordial gas, resulting in a stellar mass given by 
\begin{equation} \label{eqMstellar}
    M_\star =    M_\mathrm{vir} \mathrm{\frac{\Omega_b}{\Omega_m}}\epsilon
\end{equation}
where $\epsilon$ designates the star formation efficiency (the fraction of the baryonic mass of the halo converted into stars during a single star-formation episode). Expressing $n_\mathrm{min}$ as a function of the total stellar mass of the Pop III galaxy enables us to make comparisons with contemporary simulations predicting the number density of atomic cooling halos which has a mass according to eq.~\ref{eqMstellar}. The minimum halo mass which can host Pop III star formation is a function of redshift and spatial location and is set by the local Lyman-Werner flux as well as feedback from reionization of the local intergalactic medium. In figure~\ref{fig_nminVSz} we consider a range of star formation efficiencies $\epsilon = \{0.001,0.005,0.01,0.1\}$ in order to calculate the expected total stellar mass of atomic cooling halos across redshifts, which can then be used to estimate the minimum required number density of Pop III galaxies with that mass. Furthermore, we use a fiducial value $\epsilon = 0.01$ to calculate the masses of the atomic cooling halos simulated in \cite{2010Stiavelli} and \cite{2018Inayoshi} that goes into figures~ \ref{fig_Tele} \& \ref{fig_IMFvar}. The correct value for the star formation efficiency is indeed a matter of debate and is also important for the results of this paper -- high values produce bright objects that will be more easily detected since less magnification is required, which in turn relaxes the required minimum number density that we calculate in this paper. The opposite is of course true for low star formation efficiencies. \cite{2010Stiavelli} argues that $\epsilon \gtrsim 0.15$ is very unlikely, \cite{2020Visbal} explores a range of $\epsilon = (0.0001-0.005)$ but uses $0.001$ as their fiducial value. Some earlier work have suggested somewhat higher ranges $\epsilon = 0.01-0.1$ \citep[see e.g.,][]{2009Wise} while \cite{2018Jaacks} suggests $\epsilon = 0.05$. In \cite{2020Skinner} they suggest H$_2$ self-shielding Pop III star-forming halos typically having $\epsilon = 0.0001-0.001$. Also, \cite{2012Zackrisson} showed, using the halos simulated in \cite{2010Stiavelli,2009Trenti}, that high Pop III galaxy number densities with star formation efficiencies $\epsilon \gtrsim 0.1$ are inconsistent with the observed UV luminosity function \citep[see e.g.,][]{2012Oesch,2011Bouwens} at redshifts $z\approx 7-10$.

\subsection{Telescopes and survey parameters}
The observational facilities best suited to conduct blind searches for rare and faint objects such as lensed Pop III galaxies at high redshift are space telescopes, due to their combination of large survey area, angular resolution and high sensitivity. In this paper, we focus on the expected capabilities of the \textit{James Webb space telescope}, the \textit{Roman space telescope} and \textit{Euclid}. We emphasize that this does not mean that these are the only viable options for studying Pop III signatures in the early Universe. For example, \cite{2021Grisdale} recently demonstrated the potential for detecting He II $\lambda$1640 emission line from Pop III galaxies using the ground based \textit{ELT}/HARMONI spectrograph. However, due to the small field of view of this instrument, the success of such an approach hinges on knowing where to point the \textit{ELT}, as we have no a priori information on the location of Pop III galaxies on the sky\footnote{But see \cite{2019Johnson} for a scenario in which Pop III galaxies form in the vicinity of high-redshift quasars, which would allow clues to their preferred location.}. The selected targets would hence first have to be culled from detections in a larger-area survey of the type we consider here.

In table~\ref{surveyparams} we summarize the assumed survey parameters, based on surveys described in \citet{2019RiekeGTO,2017Marchetti,2019Akeson} and \citet{2011Laureijs}. As can be seen, \textit{Euclid} provides the largest survey areas of 15000 or 40 deg$^2$ in the wide or deep surveys respectively -- this is followed by \textit{RST} which has 2000 or 40 deg$^2$ in its wide and deep surveys respectively, although the deep spectroscopic survey covers only 12 deg$^2$. \textit{JWST}, on the other hand, only reaches survey areas of around 0.05 deg$^2$ in the surveys included. The small survey area of \textit{JWST} is greatly compensated by its depth which is considerably better than \textit{Euclid} and \textit{RST}. However, since rare, high-magnification lines of sight are also required to detect Pop III galaxies with properties similar to those seen in current simulations, it is a priori not clear which telescope will perform the best.

Considering \textit{JWST}, we can compare the assumed survey parameters to the by now established cycle 1 general observer, guaranteed time observations and early release science programs. A non-exhaustive selection of programs that are relevant for the kind of surveys suggested in this paper are given by; JADES \citep[see e.g.,][]{2017Eisenstein}, CEERS \citep{2017Finkelstein}, WDEEP \citep[reaching emission line fluxes of $\sim 10^{-18}$ erg s$^{-1}$ cm$^{-2}$ and imaging to $\mathrm{m_{AB}} \sim 30.6-30.9$;][]{2021Finkelstein}, FRESCO \citep[covering $\sim 0.017$ deg$^2$ while reaching emission line fluxes $\sim 3.3 \times 10^{-18}$ erg s$^{-1}$ cm$^{-2}$;][]{2021Oesch}, UNCOVER \citep[targeting Abell 2744 with imaging to $\mathrm{m_{AB}} \sim 29.5-30$ where emission lines can be measured for all sources with $\mathrm{m_{AB}} < 30$;][]{2021Labbe}, the WIDE MOS survey \citep[reaching $\mathrm{m_{AB}} \sim 24$ while taking spectra for objects with H$\alpha$ emission line fluxes $>10^{-17}$ erg s$^{-1}$ cm$^{-2}$ over an area of $\sim 0.1$ deg$^2$, see e.g.,][for one of the targeted fields]{2017Ferruit}. 
Note that the survey parameters taken from \cite{2019RiekeGTO} and used in table~\ref{surveyparams} summarizes imaging and spectroscopy from the JADES and CEERS programs, as such, we use these as representative parameters to asses \textit{JWST}s capabilities. However, as we see, there are multiple programs operating at different depths and over different survey areas which should in principle be summed to assess \textit{JWST}s overall likelihood of detecting pristine Pop III galaxies in any of the first cycle programs. In reality this implies slightly better prospects for detection with \textit{JWST} than the results cited in this paper, where we focus on the summarized JADES and CEERS survey presented in \cite{2019RiekeGTO}. Furthermore, several approved programs that include targets of individual cluster lenses, e.g., \citet{2017Windhorst} and \citet{2017Treu} will introduce competitive alternatives to the wider fields considered in this paper. 

\begin{table}
	\centering
	\caption{The relevant survey parameters for the three telescopes considered (5$\sigma$ point source detection limits). The \textit{JWST} numbers for NIRCam and the deep NIRSpec (0.013 deg$^2$ survey with $\sim 28$ hr exposures) are quoted from planned surveys \citep{2019RiekeGTO}, while the NIRSpec values for a single field of view $\sim 100$ hr NIRSpec exposure (very deep) are estimated using the exposure time calculator. We include two different setups with \textit{RST} \citep{2019Akeson} and \textit{Euclid} \citep{2011Laureijs,2017Marchetti} in order to highlight the utility of using either wide surveys or deep surveys.}
	\begin{tabular}{c c c c} 
		\hline \hline 
		Telescope & Survey area &  Photo. depth & Spec. depth \\
		 & (deg$^2$) & (AB mag) & (erg s$^{-1}$ cm$^{-2}$)   \\
		\hline
		 deep NIRCam/ & 0.013 & 30.6  & $2.9\times 10^{-19}$  \\
		NIRSpec &  &  &  \\
		 med. NIRCam & 0.053 & 29.7  & --  \\
		 very deep NIRSpec & 0.0034 & --  & $1.3\times 10^{-19}$  \\
		 deep \textit{RST} (photo.) & 40 & 29 & -- \\
		 deep \textit{RST} (spec.) & 12 & -- & $1\times 10^{-17}$ \\
		 wide \textit{RST} & 2000 & 27 & $1\times 10^{-16}$ \\
		deep \textit{Euclid} & 40 & 26 & $5\times 10^{-17}$ \\
		wide \textit{Euclid} & 15000 & 24 & $3\times 10^{-16}$ \\
		\hline \hline
	\end{tabular}
    \label{surveyparams}
\end{table}

\subsection{Minimum comoving density for the detection of lensed Pop III galaxies}
We derive the comoving volumes spanned by the survey areas of the different telescopes by integrating the comoving differential volume element over the considered redshift and survey area. The depth in redshift of the comoving volumes is set by the typical redshift range ($\Delta z = 1$) in which galaxies are picked using the Lyman dropout technique using broadband filters. The comoving volume, defined here as $V_c$, is a function of the cosmological model used, the redshift ($z$), redshift bin-size ($\Delta z$) and the survey area (units of square degrees, defined as $A$). In a flat Friedmann-Lemaitre-Robinson-Walker spacetime, the comoving volume is given by:
\begin{equation} \label{eqVc}
      V_c = \frac{4 \pi}{3}\frac{A}{41253}\left[D_c(z+\Delta z/2)^3 -D_c(z-\Delta z/2)^3\right]
\end{equation}
where $D_c(z\pm \Delta z/2)$ is the comoving distance at redshift $z\pm \Delta z/2$. Given the comoving survey volume $V_{\mathrm{c}}(A, z, \Delta z)$  probed at each redshift, we can calculate the minimum comoving number density of chemically pristine Pop III galaxies ($n\mathrm{_{min}}$ required in order to detect at least one Pop III galaxy candidate in one single survey, by applying the constraint
\begin{equation} \label{nmineq}
    n\mathrm{_{min}}=\frac{1}{P(> \mu_{ \mathrm{min}}, z) \times V_{\mathrm{c}}(A, z, \Delta z)}.
\end{equation}

A time window ($t_\mathrm{Pop III} = 10$ Myr) is later applied to the simulations where necessary, in order to make sure that we calculate the number of chemically pristine Pop III galaxies inside the comoving volume, that is, galaxies that would be identifiable as Pop III before losing its unique diagnostic observational signatures. 

Notice the dependency on $\mathrm{\mu_{min}}$, which in turn is a function of the emission line flux $F_\mathrm{HeII \,1640}$ for spectroscopy, or, the continuum flux $F_\mathrm{UV \,1500}$ in the case of photometry. The minimum magnification $\mathrm{\mu_{min}}$ determines the probability for gravitational lensing to boost the intrinsic flux of a galaxy above the detection threshold of the considered telescope/survey. This is calculated through the magnification probability function $P(> \mu, z)$ -- which determines the probability of getting a magnification greater than $\mu$. 

The logic goes as follows; for a given comoving number density ($n$) we can find the total number of objects inside a comoving volume $V_c$ by multiplying the two. Whether we can detect e.g., the He II $\lambda$1640 emission line from objects inside this volume depends on the telescope's detection limit, the emission line flux ($F_{\mathrm{HeII \,1640}}$) and the magnification ($\mu$) from gravitational lensing. The required magnification ($\mu_{\mathrm{min}}$) is clearly a decisive parameter when it comes to estimating $n\mathrm{_{min}}$. We should therefore ask ourselves what minimum magnifications we would require to detect a Pop III galaxy, either spectroscopically, or photometrically. In the case of spectroscopy, the answer is given by:
\begin{equation} \label{eqmuSpec}
    \mu_{\mathrm{min}} = \frac{4\pi D_L(z)^2 F\mathrm{_{limit}}}{\mathcal{L}_\mathrm{HeII \,1640}}\frac{\tau}{M_{\star}}
\end{equation}

where $F_\mathrm{limit}$ is the spectroscopic detection limit of the telescope considered (see table~\ref{surveyparams}), $D_L(z)$ is the luminosity distance to the galaxy, $\mathcal{L}_\mathrm{HeII \,1640}$ is the He II $\lambda$1640 emission line luminosity per solar mass per year, $M_{\star}$ the total stellar mass of the galaxy and $\tau$ is the age of the galaxy after onset of star formation in the galaxy. Note that what we have above is simply a ratio between the telescope flux detection limit and the emission line flux, scaled up or down with the average (constant) star formation rate; SFR$^{-1}=\frac{\tau}{M_\star}$.

In the case of photometry, we use the rest frame luminosity density at $1500 \, \ang$ to estimate the detectability of Pop III galaxies. Using AB magnitudes and a factor $(1+z)$ for the K-correction, we have the following:
\begin{equation} \label{eqmuPhoto}
     \mu_{\mathrm{min}} = \frac{c}{(1+z)\lambda_0^2}4\pi D_L(z)^2 10^{-(m_\mathrm{limit}+48.6-25)/2.5} \frac{\tau}{M_{\star}} \frac{1}{\mathcal{L}_\mathrm{UV \, 1500}}
\end{equation}
where instead we now have the dependence on the photometric detection limit ($m_\mathrm{limit}$ -- in AB magnitudes), the rest frame wavelength $\lambda_0$ (in our case $\lambda_0 =1500\, \ang$) and $\mathcal{L}_\mathrm{UV \, 1500}$ which now designates the luminosity density at 1500 $\ang$ (units of erg s$^{-1}$ $\ang^{-1}$), normalized to $1M_\odot$ yr$^{-1}$, and $c$ designates the speed of light.

The fiducial star forming age at which we extract the spectrum of the Pop III galaxies is given by $\tau=10$ Myr. In reality the star forming age of the observed galaxy would form a distribution of values that for the sake of observational availability of the object would have to be bounded by $\tau < t_\mathrm{popIII} = 10$ Myr. Since the luminosity of the galaxy is dependant on the star forming age (the UV/line luminosity generally ramps up with time quickly as star formation has commenced) we could in principle observe the galaxy at a very early stage as it has not yet reached its equilibrium luminosity -- making it somewhat harder to detect. For example, between 1 and 2 Myr, the luminosity in the He II $\lambda$1640 emission increases by $\sim 22$ per cent while the continuum luminosity at 1500 $\ang$ increases by a factor of $\sim 2.3$ in the same time span. For the Pop III, top-heavy IMFs considered here, we find that after $\tau = 2$ Myr the galaxy has reached a stable luminosity after which it does not change significantly and therefore defines an age from which the luminosity that the galaxy has will be retained for a very long period, given a constant star formation rate -- making $\tau = 10$ Myr a suitable fiducial value. We will discuss the effects of looking at lower star forming ages further in section~\ref{ImpactofAge}

\subsection{Detection probability}
The number density limit in eq.~\ref{nmineq} is valid for a certain probability $P_\mathrm{det}$ to end up with at least one Pop III galaxy above the detection threshold. The probability of lifting one pristine Pop III galaxy, out of $N$ possible, for a given redshift, above the detection limit is given by; $P_\mathrm{det} = 1-[1-P(> \mu\mathrm{_{min}})]^{N}$, where $N$ is the number of pristine galaxies observable during any time window $t_\mathrm{Pop III}$ within the $\Delta t$ (the cosmic time spanned by the surveyed volume) which require the magnification $\mu\mathrm{_{min}}$ for detection. 

Demanding a specific probability to lift one galaxy above the detection limit therefore sets a relation between $N$ and $P$ for the given required magnification. For example, to get a 95 per cent probability ($P_\mathrm{det} =0.95$, i.e., $2\sigma$ level) of detecting the He II $\lambda$1640 emission line from a single galaxy with total stellar mass $M_{\star}\approx 4.5\times10^4\, M_\odot$ using \textit{JWST} with the depth $1.3\times 10^{-19}$ erg s$^{-1}$ cm$^{-2}$, at $z=10$, our computational machinery indicates that we need a magnification of $\mu\mathrm{_{min}} \approx 64$, which has a probability $P(>\mu\mathrm{_{min}})\approx 2\times 10^{-6}$ of occurring. This results in $N\approx \frac{\ln(1-P_\mathrm{det})}{\ln(1-P(> \mu\mathrm{_{min}}))} $ -- implying that we need approximately 1.4 million pristine galaxies (i.e., with a star forming age $\tau < 10$ Myr) inside the whole survey volume considered by \textit{JWST}, in order to make a single detection of a lensed galaxy.
 
The relation in eq.~$\ref{nmineq}$ is not as stringent, instead having a scaling $N = \frac{1}{P}$, leading to a requirement of $N \approx$ 460 000 pristine galaxies in the survey volume considered -- about a factor of 3 less than that required for the 2$\sigma$ level. We therefore consider eq.~$\ref{nmineq}$ as a lower bound on the minimum comoving number densities presented in Section~\ref{results}. 

\section{Results} \label{results}
In Figure~\ref{fig_Tele} we show that all included telescopes require more or less identical minimum required number densities for spectroscopic detection of the He II 1640$\lambda$ Å emission line. However, due to limitations in the maximum magnifications $(\mu_\mathrm{max} \approx 1000)$ that are applicable, \textit{JWST} is the only telescope that has the required sensitivity to detect atomic cooling halos with lower star formation efficiencies (e.g., $\epsilon = 0.001$) up to high redshifts $(z\lesssim 11.5)$. We find that \textit{JWST} does better than the other telescopes for the most massive Pop III galaxies $(M_\star \gtrsim 10^6\ M_\odot)$. We also find that the \textit{Euclid} wide survey perform better than its deep survey when it comes to detecting the He II 1640$\lambda$ Å emission line. However, the magnification limit ($\mu_\mathrm{max}$) at $z=10$ suggests that the wide \textit{Euclid} survey can only be applied to total stellar masses $\gtrsim 6.7 \times 10^6\, \Msol$, which is notably larger than predicted by the simulations included. \textit{RST}, on the other hand, can reach total stellar masses of $M_\star \approx 2 \times 10^5\, \Msol$ in its deep survey without breaching the magnification limit at $z=10$. In the case of photometric detection of the UV-continuum at 1500 Å at $z=10$, the best telescope is \textit{RST} for all the total Pop III galaxy stellar masses. However, we find a notable area around $M_\star \gtrsim 7\times 10^5 \ M_\odot$ where \textit{JWST} can be competitive in deep and medium-deep exposures. The photometric wide \textit{RST} survey can reach Pop III galaxies forming in atomic cooling halos with star formation efficiency $\epsilon = 0.01$ for $z\lesssim 11.5$, without breaching the magnification limit -- for \textit{Euclid} on the other hand, the deep survey is necessary, which reaches $z\lesssim 8$ without breaching the magnification limit.  

The individual lines in figure~\ref{fig_Tele} display segments with different slope in the different mass ranges, depending on the survey. The flat, horizontal part of the lines indicates the region where the telescope does not require gravitational lensing to detect the galaxy. The location that the horizontal line has on the y-axis is determined by the survey area of the telescope. The location where the lines bend upwards is set by the depth of the telescope and therefore also marks the mass below which gravitational lensing is required -- and as we move to the left along the lines, the magnification $(\mu)$ becomes larger. The slope of the line is steeper to begin with but levels off at a certain point -- this occurs when the required magnification reaches $\mu \approx 10$ where the magnification probability function enters the $P(\mu)\propto \mu^{-2}$ regime, as is described by the polynomial fit in eq.~\ref{eqPfit}. We separate the figures into a ''detectable'' region and a ''undetectable'' region, where the separation is determined by the lines belonging to the different surveys. Any simulation that predicts Pop III galaxy number densities above these lines implies that there is a good chance for detection -- placing it in the detectable region. Likewise, if the predicted number density is below the lines, prospects for detection are poor -- therefore placing it in the undetectable region.

Apart from the telescope survey area and depth, the relevant parameters for the minimum Pop III galaxy number density for detection are; the age of Pop III burst ($\tau$), the Pop III IMF and the redshift of the observed galaxy. Varying the survey area ($A$) will shift the lines in figures~\ref{fig_Tele} \& \ref{fig_IMFvar} up (if $A$ is decreased) or down (if $A$ is increased), while the depth of the telescope shift the lines to the left (for higher depth, i.e. higher sensitivity) or right (for lower depth). The survey volume scales linearly with survey area (eq.~\ref{eqVc}), implying that $n_\mathrm{_{min}}$ scales inversely with $A$, while the detection limit of the telescope enters quadratically into the magnification probability function (see eq.~\ref{eqPfit}) due to its dependence on $\mu\mathrm{_{min}}$  (see eq.~\ref{eqmuSpec} \& \ref{eqmuPhoto}).

\subsection{The detectability of simulated galaxies} \label{Detectability}
Looking again at figure~\ref{fig_Tele}a indicates that, at $z=10$, the prospects for spectroscopic detection of the He II $\lambda$1640 $\ang$ emission line in the surveys included in this paper are not ideal. For example, even the simulations in \cite{2010Stiavelli} suggests number densities that are 1-2 orders of magnitude lower than what is required from our calculations (depending on which of the two total stellar masses that is considered). We keep in mind though that this prediction hinges on the value of the star formation efficiency, which in figure~\ref{fig_Tele} is set to 0.01 for the \cite{2010Stiavelli} and \citet{2018Inayoshi} simulations. The calculations also show that all the included surveys require very similar minimum number densities for detection, however, \textit{JWST} is the only telescope that does not breach the maximum allowed magnification at this redshift. When it comes to photometry, the prospects for detection are quite good when comparing to the predictions of the simulations. \textit{RST} and \textit{Euclid} surveys are predicted to detect Pop III galaxies using photometry, given the number densities suggested by \cite{2010Stiavelli}. We also see that both the wide and deep \textit{RST} surveys should detect the more massive galaxies predicted by \cite{2020Visbal} in their high efficiency ($\epsilon=0.005$) simulations while the galaxies simulated in their fiducial model ($\epsilon = 0.001$) has number densities that are at the very edge minimum requirements. For the \citet{2010Stiavelli} simulations we also find that \textit{JWST} may be able to detect about one Pop III galaxy if its total stellar mass is $\gtrsim 10^5 \ \Msol$ . Looking at figure~\ref{fig_nminVSz}, we can see the redshift evolution of the minimum required comoving number density for spectroscopic detection of the He II $\lambda$1640 $\ang$ emission line. Here, we can once again see that our fiducial $\epsilon = 0.01$ model requires number densities that are higher than the predictions from simulations. However, this figure suggests that if the star formation efficiency can be pushed up to $\epsilon = 0.1$, \cite{2010Stiavelli} predicts number densities that are sufficient at $z\sim 10$ -- but such high star formation efficiencies together with their predictions for the number density of atomic cooling halos were shown to be inconsistent with the UV luminosity functions at $z\approx 7-10$ by \cite{2012Zackrisson}. Note that \cite{2020Visbal} has a well defined star formation efficiency in their models which has an effect on the predicted number densities. The star formation efficiency is linked to the intensity of the Lyman-Werner background which sets the minimum mass where Pop III stars can form -- therefore requiring additional simulations in order to accurately compare with the higher star formation efficiency.

If the star formation efficiency is indeed lower than $\epsilon \approx 0.1$, we see that spectroscopic detection of a Pop III galaxy is very improbable in the surveys included here. A way to improve the prospects for detection is therefore to consider even wider and/or deeper surveys. As stated, increasing the depth of the telescope has a larger impact than increasing the survey area -- but going deeper unfortunately demands significantly higher exposure times.

\subsection{The impact of the Pop III stellar IMF}\label{ImpactofIMF}
\begin{figure} 
  \centering
    \includegraphics[width=\columnwidth]{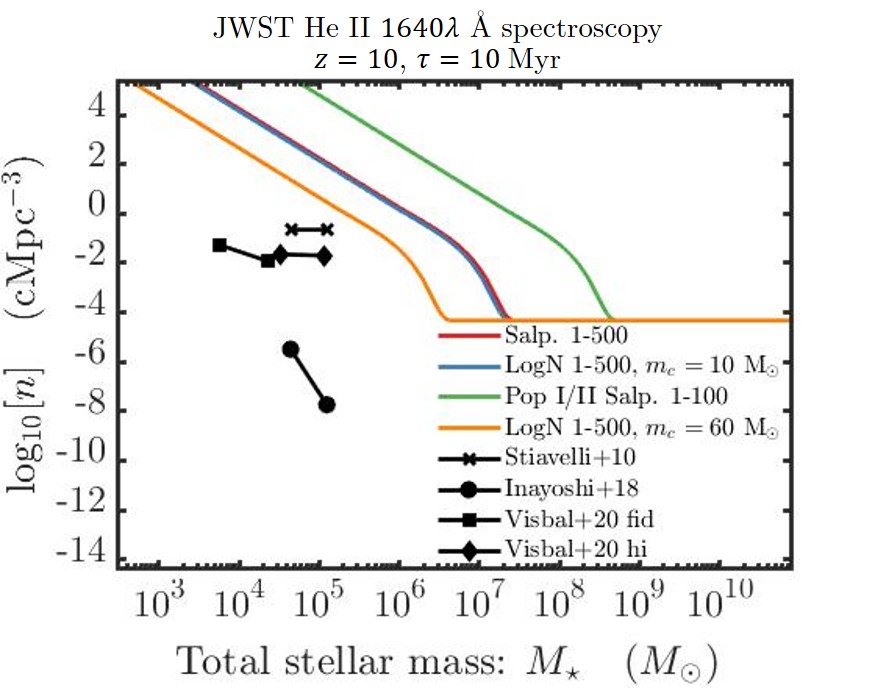}
    \caption{The impact of varying the Pop III IMF on the minimum required number density for the very deep NIRSpec survey mode. The baseline (the flat, lower part of the lines) stay the same irrespective of the IMF since it only affects the luminosity of the source, resulting in a horizontal shift. Note that the lognormal, $m_c = 10 \, M_\odot$, $\sigma = 1$, in the mass range $1-500\, M_\odot$ (blue line) and Salpeter $1-500\, M_\odot$ (red line) IMFs are almost completely overlapping. We have also included a standard, solar metallicity Pop I/II Salpeter IMF (green line) as a comparison. From this result we conclude that the lognormal IMF , $m_c = 60 \, M_\odot$, $\sigma = 1$, in the mass range $1-500\, M_\odot$ that is used in fig.~\ref{fig_Tele} is indeed the most optimistic -- lowering the Pop III stellar mass required for detection by an order of magnitude when compared to the less top-heavy IMFs. The simulations included are the same as in figure~\ref{fig_Tele}.}
     \label{fig_IMFvar}
  \end{figure}

\begin{figure} 
  \centering
    \includegraphics[width=\columnwidth]{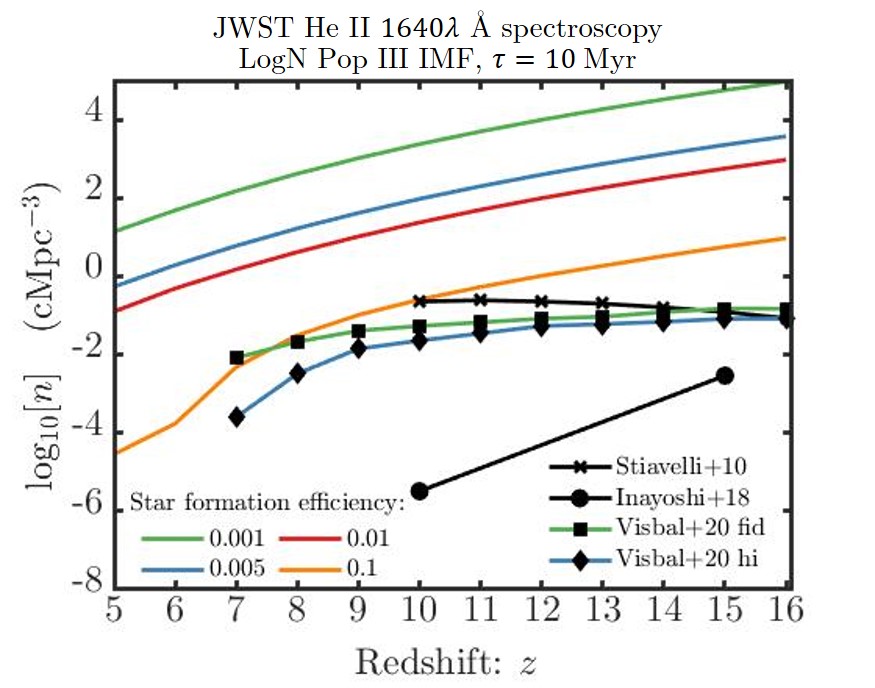}
    \caption{The minimum number density $(n_\mathrm{{min}})$ required for spectroscopic detection of the He II $\lambda$1640 $\ang$ emission line in suitable halos (atomic cooling) using the very deep \textit{JWST} survey, as a function of redshift. The lines correspond to different star formation efficiencies ($\epsilon$) in the range 0.001 - 0.1. The IMF used is the fiducial lognormal top-heavy, $m_c=60 \, M_\odot$, $\sigma = 1$ in the mass range $1-500 \, M_\odot$. The virial mass of these atomic cooling halos ($T_\mathrm{vir}=10^4$ K) are redshift dependent which, given a star formation efficiency, implies a range of total stellar masses for each line in the figure. The virial mass range is; $M_\mathrm{vir} \approx 1.5 \times 10^7 - 7 \times 10^7 \, M_\odot$, increasing from high redshift to low redshift. The total stellar mass for the plotted lines are then given by  $M_\star=M_{\mathrm{vir}} \mathrm{\frac{\Omega_b}{\Omega_m} \epsilon} \approx M_{\mathrm{vir}} 0.15 \epsilon$ (cf. eq.~\ref{eqMvir} \& \ref{eqMstellar}). The included simulations (black markers) provide the number densities of atomic cooling halos. Note that the \protect\cite{2020Visbal} simulations here are predictions for $\epsilon= \{0.001,0.005\}$ -- therefore the color on the lines for these simulations have been matched with the color of the minimum required number density with the corresponding star formation efficiency. Furthermore, the total Pop III stellar masses predicted in \protect\cite{2020Visbal} vary slightly from the mass at the atomic cooling limit at the higher redshifts as star formation is then allowed in halos with lower masses. Notice the different behaviour for the $\epsilon=0.1$ line which produces a galaxy which is massive enough (i.e., bright enough) for it to not require gravitational lensing at the lowest redshift, but begin to require magnification around $z=6.5$. All the other star formation efficiencies require gravitational lensing throughout the redshift range. The $\epsilon=0.1$ line is the only realization which produces Pop III galaxies that are even close to some of the simulations. This plot can also be applied to \textit{RST} and \textit{Euclid} as long as the total stellar mass of the Pop III galaxy is not too low -- which restricts the redshifts at which it is valid. At $\epsilon = 0.01$, the deep \textit{RST} survey have enough sensitivity to spectroscopically detect atomic cooling halos without breaching $M_{\star,\mathrm{min}}$ for $z \lesssim 7$, while deep \textit{Euclid} surveys cannot even reach $z \sim 5$ -- pushing the star formation efficiency to $\epsilon = 0.1$ puts these limits at $z \lesssim 12$ and $z \lesssim 8$ for deep \textit{RST} and \textit{Euclid} surveys respectively. The very deep NIRSpec survey is only affected by this limit in the case of $\epsilon = 0.001$, where we are restricted to $z \lesssim 11.5$.}
    \label{fig_nminVSz}
  \end{figure}

\begin{figure} 
  \centering
    \includegraphics[width=\columnwidth]{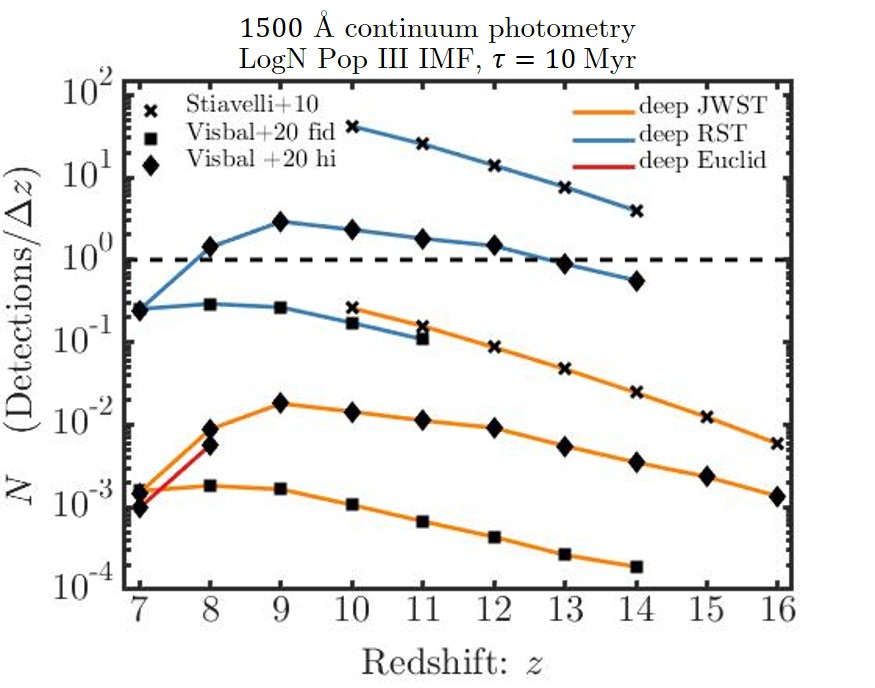}
    \caption{The expected number ($N$) of photometrically detected Pop III galaxies per redshift bin ($\Delta z = 1$) as a function of redshift for the different survey and simulation combinations. The redshift range applicable to the different surveys and simulations are restricted by the maximum allowed magnification $\mu_\mathrm{max}=1000$ as well as the limit at which the $1500 \, \ang$ continuum redshifts out of the surveys reddest filter. The restrictions can be notably seen for \textit{Euclid} which is only applicable for the lowest redshift ranges combined with the \citet{2020Visbal} high efficiency model due to magnifications limits. The deep \textit{RST} survey is restricted to $z\lesssim 11$ due to magnification limits when combined with the fiducial \citet{2020Visbal} simulations, while for the other simulations, the $1500 \, \ang$ continuum redshifts out of the \textit{RST} filters for $z\gtrsim 14$. The deep \textit{JWST} survey is only restricted by magnification limitations when combined with the \citet{2020Visbal} fiducial simulations where it reaches $z\lesssim 14$. The different surveys are identified by the color on the lines while the combination of that particular survey with a simulation is indicated by the respective black marker (cross, square or diamond). The dashed black line illustrates which survey and simulation combinations that are expected to detect at least one Pop III galaxy per redshift bin. As revealed, \textit{RST} is the only telescope that is predicted to detect at least one Pop III galaxy per redshift bin over a considerable redshift range when combined with the \citet{2010Stiavelli} or the high efficiency \citet{2020Visbal} simulations.}
    \label{fig_Nvsz}
  \end{figure}

 As mentioned earlier, we use a fiducial lognormal IMF in the mass range $1-500 \,M_\odot$ for our predictions. The strength of the He II $\lambda$1640 $\ang$ emission line is affected significantly by the IMF considered. This is due to the correlation between the abundance of high-mass stars $(\gtrsim 50 \,M_\odot)$ and the amount of hard ionizing flux. A standard Salpeter IMF in the mass range $\sim 1-100 \,M_\odot$ simply won't produce the required ionizing flux needed to detect the He II $\lambda$1640 $\ang$ emission line without very strong magnification. As figure~\ref{fig_IMFvar} reveals, the same goes for other Pop III IMFs which have suppressed high-mass tails, resulting in low He II $\lambda$1640 $\ang$ fluxes and therefore poor prospects for detection. Similarly, the IMF impacts the continuum level at $1500 \ \ang$, which affects the detectability of the galaxies in photometric surveys. The results of figure~\ref{fig_Tele}, \ref{fig_nminVSz} \& \ref{fig_Nvsz} were extracted for the lognormal IMF with characteristic mass $m_c = 60 \, M_\odot$ and width $ \sigma = 1$, and a mass range $1-500\,M_\odot$. In the case that nature in reality provides us with a less top-heavy IMF, the requirements on the minimum number density increases by at least an order of magnitude for the other, less extreme models included here.

\subsection{The impact of the Pop III galaxy age} \label{ImpactofAge}
For the constant star formation rate considered in this paper, the age of the Pop III burst is important early after star formation has initiated, as the He$^+$-ionizing flux grows rapidly at the onset of star formation while stabilizing after $\sim 2$ Myr to remain more or less constant throughout the considered time span. The actual variations in luminosity with age is only at the level of a few per cent in the range of ages 3 - 10 Myr -- for the continuum level at $1500 \, \ang$ as well as the He II $\lambda$1640 $\ang$ emission line luminosity. Therefore, once the initial $\sim 2$ Myr burst of star formation has completed, the impact of the star forming age does not significantly affect the detectability of Pop III galaxies. However, as the Pop III galaxy total stellar masses predicted by simulations provide the mass that has been converted into stars before the galaxy proceeds into chemically enriched star formation, the minimum required number density that we predict for a given Pop III galaxy total stellar mass is therefore specifically calculated for galaxies that form their stars over the 10 Myr period that we use as a fiducial age. Performing our calculations for the same total stellar mass but at lower star forming age implies a higher SFR, which would render the galaxies brighter and thus more easily detectable. In order to investigate how this affects the detectability of Pop III galaxies, we also performed calculations at a lower star forming age. Looking at shorter star forming ages also requires the included simulations to be scaled accordingly in order to not overestimate the number of pristine Pop III galaxies. The scaling of simulated number densities with age is linear while the effects of age on our calculated minimum required number densities scales quadratically (this can be seen by combining eq.~\ref{eqPfit}, \ref{nmineq} and \ref{eqmuSpec}). The slight variation of the luminosity with age for a constant SFR also enters quadratically into the minimum number density calculation, but as mentioned above, this effect is very small. For example, performing our calculations at a star forming age of 3 Myr instead of 10 Myr lowers the minimum number density required to detect one object by $\sim$ 1 order of magnitude while simultaneously lowering the number density predicted by simulations by $0.5$ orders of magnitude -- effectively reducing the gap between the simulations and our minimum required number density with $\sim 0.5$ orders of magnitude.

\subsection{The redshift dependence} \label{theredshiftdepedence}
The dependence on redshift are two-fold in these calculations. First off, a more distant galaxy of a given mass (and therefore integrated luminosity) are simply dimmed according to the luminosity distance relation. Increasing the redshift therefore shifts the detection delimiter lines in figures~\ref{fig_Tele} and \ref{fig_IMFvar} to the right. Secondly, in our calculation of the comoving survey volume, the fixed redshift bin size of $\Delta z$ implies that we are probing different comoving volumes at each redshift. Due to the slow increase of comoving distance with redshift, as we go to higher $z$, the bin size of $\Delta z$ covers a smaller depth. The increase of transverse comoving length spanned by the telescope field of view does not compensate for this, implying that the comoving survey volume ($V_c$) decreases with redshift -- which in turn increases $n_\mathrm{_{min}}$ -- shifting the lines in figures~\ref{fig_Tele}-\ref{fig_IMFvar} upward. Typically, surveys will cover more than a single such $\Delta z$ slice of spacetime which would imply that any of the surveys included here may observe Pop III galaxies at several redshifts provided that the number densities are high enough. This would in principle lower the minimum required number density at each redshift if we are content with detecting only one Pop III galaxy at any of the probed redshifts. This will only affect our results moderately, but the effect increases depending on how many individual redshift bins that are covered in the observations. Increasing/decreasing $\Delta z$ results in larger/smaller volumes being probed at each redshift.

The redshift that is the most promising for detection depends on the star formation efficiency. Looking at figure~\ref{fig_nminVSz} we see that for the high star formation efficiency of 0.1, \cite{2010Stiavelli} produce the required number density at redshift $z\approx 10$. In the more realistic scenario of lower star formation efficiencies, simulations are closest at, or around, the peak of Pop III star formation at $z\lesssim 10$, which motivates our choice of fiducial redshift for the other figures. Lower redshifts are however of great interest if the simulated Pop III number densities can remain at a sufficiently high level.

In figure~\ref{fig_Nvsz} we display the expected number of photometric Pop III galaxy detections per redshift bin ($\Delta z =1$) as a function of redshift. By applying the different surveys to the predicted number density from the simulations, we can estimate how many Pop III galaxies we expect to be detected. Once again, we assume the lognormal top-heavy IMF and 10 Myr star forming age. Furthermore, we assume the star formation efficiency $\epsilon = 0.01$ for the Pop III galaxies forming in atomic cooling halos predicted by \citet{2010Stiavelli}, while the \citet{2020Visbal} simulations provide specific total stellar masses (as given by their high-efficiency and fiducial model). Notice that there are two different total stellar masses available for each simulation -- we use the lower as it is representative of the typical atomic cooling halo in which Pop III galaxies form. The \citet{2018Inayoshi} simulations have not been included due to their very low number densities -- far from producing any detections. Furthermore, we have only plotted surveys which do not breach the maximum magnification limit and have also excluded surveys in which the 1500 $\ang$ continuum redshifts out of the telescopes available filters. As revealed by the figure, the only survey and simulation combination that predicts detections of Pop III galaxies across the whole considered redshift range, is the deep \textit{RST} survey combined with the \citet{2010Stiavelli} simulation while the high efficiency \citet{2020Visbal} simulations provide detection between redshifts $z\approx 8-13$ in the same survey.

Due to the nature of conducting high redshift surveys, where the telescopes scan a range of redshifts as one continuously reaches for fainter objects, we can estimate the expected number of detections across all redshifts where we have access to simulations. This reveals that the deep \textit{JWST} survey would detect $\sim 0.6$ Pop III galaxies if combined with the \citet{2010Stiavelli} simulations (e.g., 2 times more survey area is required in order to produce $\sim$1 detection). The deep \textit{RST} survey produces a considerable number, about 92, of detections if combined with the \citet{2010Stiavelli} simulations, a total of $\sim 12$ if combined with \citet{2020Visbal} high efficiency model or $\sim 1$ if combined with the fiducial \citet{2020Visbal} model. The wide \textit{Euclid} photometric survey cannot provide detections without breaching the maximum magnification limit set in this paper.

\section{Discussion} \label{discussion}
Considering the results of this paper (based on the surveys in table \ref{surveyparams}), the prospects for spectroscopic confirmation of the true Pop III nature of high-redshift galaxies using the He II $\lambda 1640$ emission line appear to remain grim. However, in this paper we have set a stringent requirement on the kind of stellar population, namely that it is truly pristine, such that it contains absolutely no metals at all and simultaneously shows a strong He II $\lambda 1640$ line. When searching the skies in future high-redshift surveys, we will potentially find many cases of so-called hybrid Pop III galaxies, in which Pop III stars co-exist with metal-enriched stars \citep{Sarmento18,Sarmento19}. These objects are still very interesting when it comes to understanding the formation of the first truly pristine galaxies as it will bring deeper insight into the timescales related to metal pollution and the feedback effects present at the onset of star formation and put potential constraints on the masses of Pop III stars.

Furthermore, in the estimations of the gravitational lensing probabilities used in this paper, we based our calculations on surveys performed along random sightlines. This generally provides probabilities that are much lower than what one would expect when targeting known galaxy clusters acting as lenses. In these so called cluster lenses one has enhanced probabilities of achieving the very highest magnifications which may very well be the cases which turn out to provide us with the first spectroscopic detection of a truly pristine Pop III galaxy (see section below). When utilizing gravitational lensing as a necessary condition for detection one must also keep in mind that there are complicating factors that may compromise observations due to the presence of so called blends -- i.e., the contamination of flux from other objects, e.g., galaxies at other redshifts along the line of sight, for instance galaxies in a galaxy cluster responsible for the lensing of the Pop III object. This is a phenomena that may, in some cases, render observations that meet all the other required criteria for detection useless.    

\subsection{Comparison to cluster lensing surveys}
In this paper we have focused on the utility of blank fields (randomly located in the sky) that due to their large survey areas could be able to detect a considerable number of gravitationally lensed Pop III galaxies. The large survey areas considered in this paper stands in stark contrast to the very narrow fields relevant when targeting specifically selected cluster lenses that provide high probabilities for achieving very high magnifications. We can make a rough estimate on the capability of individual cluster lenses for the detection of Pop III galaxies compared to blank fields by taking the magnification data assembled in \citet{2012Zackrisson} that provide the area on the sky over which high $(\mu>10)$ magnifications are attained when targeting individual cluster lenses. Given the brightness of the Pop III galaxies considered in this paper, and the detection limit of the telescope in question, we can calculate the minimum required magnification and therefore also the area over which this magnification is realised. Applying this area to the minimum required number density of Pop III galaxies that we calculated for our randomly oriented wide field surveys, we can make order of magnitude estimates on whether a cluster lens or a wide field survey performs better. We consider the approved early release science program by \citet{2017Treu} and the guaranteed observer time program by \citet{2017Windhorst} which target several known cluster lenses with \textit{JWST} at deep/medium sensitivity (detection limits around $\mathrm{m_{AB}}\sim 28.5-29$ in photometry, and $\sim 3.5\times10^{-19}$ erg s$^{-1}$ cm$^{-2}$ in spectroscopy) as representative surveys to compare our calculations with.

For spectroscopy on the He II $\lambda$1640 $\ang$ emission line, we found that our wide field surveys would require a minimum number density of $\sim 26$ cMpc$^{-3}$ in order to detect a single Pop III galaxy of mass $4.4\times10^4 \ \Msol$ (the total stellar mass of an atomic cooling halo with $\epsilon = 0.01$), at $z=10$. In order to detect such a Pop III galaxy with a sensitivity of $3.5\times10^{-19}$ erg s$^{-1}$ cm$^{-2}$ one requires $\mu \sim 178$. This magnification is attained, on average, over an area of $\sim 0.082$ arcmin$^2$ in a typical cluster lens. This area maps to a comoving volume of $\sim 0.79$ cMpc$^{3}$ in the source frame of potential Pop III galaxies at $z=10$. Applying this volume to the number density of $\sim 26$ cMpc$^{-3}$ reveals that a single cluster lens is about 20 times better at spectroscopically detecting Pop III galaxies than the wide field surveys considered in this paper. Taking the number density of Pop III galaxies with age $\leq 10$ Myr that is predicted by simulations, given by $\sim 0.23$ cMpc$^{-3}$ at $z=10$ for the \citet{2010Stiavelli} model, one finds that a total of $\sim 6$ cluster lenses has to be targeted in order to detect one Pop III galaxy with spectroscopy. When looking at photometry on the other hand, the scenario is different. The survey with the best performance for photometric detection of the continuum at 1500 $\ang$ that we found in this paper was the wide \textit{RST} survey (at $z=10$) which required a minimum number density of $\sim 0.004$ cMpc$^{-3}$. Performing a similar calculation for photometry shows that a cluster lens observation operating at a depth of $\mathrm{m_{AB}}\sim 28.5$ (requiring $\mu \sim 172$ over $0.085$ arcmin$^2$) would detect a factor of several hundreds fewer Pop III galaxies than the wide \textit{RST} survey. Figure~\ref{fig_Nvsz} reveals that the wide \textit{RST} survey would detect $\sim 40$ Pop III galaxies at $z=10$, given the number densities predicted by \citet{2010Stiavelli}. This can be compared to cluster lensing which would rather need to target $\sim 5$ cluster lenses in order to photometrically detect one single Pop III galaxy at $z=10$, for the given depth of the telescope. Pushing the telescope depth to $\mathrm{m_{AB}}\sim 29.5$ we crudely estimate that a single cluster lens could detect at least one Pop III galaxy.

These estimates showcase the utility of wide photometric surveys which manages to reach the depth required to detect Pop III galaxies with the help of gravitational lensing while simultaneously covering very large areas on the sky. The benefit of targeting a cluster lens, where the probabilities for high magnifications are better, is greatly counteracted by the large relative difference between the areas covered by a wide field survey and a cluster lensing survey -- therefore resulting in better detection prospects for the wide field surveys discussed in this paper. For spectroscopy, which inherently requires very deep exposures in order to detect indicative emission lines in high-z Pop III galaxies, the benefit of frequently achieving strong magnifications in cluster lenses results in very advantageous detection prospects -- at a relatively small cost in survey area.

\subsection{Comparison to previous studies}
Previous work has showed that we are not limited by the areal number density of minihalos (the potential hosts of Pop III galaxies) in the field of view of typical telescopes \cite[e.g.,][]{2020Schauer} -- which continue to show that an \textit{ultimately large telescope} (\textit{ULT}) reaching AB magnitudes of $\sim 39$ would be able to detect even the very first Pop III stars, without gravitational lensing.

\cite{2020LiuBromm} calculate the number of Pop III galaxies per 10 arcmin$^2$ with total stellar mass $M_\star = 10^3 \, \Msol$ as a function of redshift for a variety of metal-mixing models. At $z=10$, they predict numbers of such Pop III galaxies as large as $4\times 10^3 - 2 \times 10^4$ per 10 arcmin$^2$, ranging from their pessimistic to their optimistic model. In our calculations, including the effects of gravitational lensing, we find that the minimum number of Pop III galaxies per 10 arcmin$^2$ required for $\sim$ one detection of a $M_\star = 10^3 \ \Msol$ object with the best performing surveys -- the wide and deep \textit{RST} photometric surveys -- is given by $\sim 3.7\times 10^4$ per 10 arcmin$^2$. This once again points out the importance of investigating whether high efficiency star formation really is ruled out. For example, considering Pop III galaxies with total stellar masses a factor of 10 higher, i.e., $M_\star = 10^4\, \Msol$ puts us at a required minimum of $\sim 3.7\times 10^2$ Pop III galaxies per 10 arcmin$^2$, at $z=10$, for the wide and deep \textit{RST} survey. Even more massive Pop III galaxies -- as high as $M_\star \approx 8 \times 10^4 - 2.5\times 10^6 \, \Msol$ -- have been suggested by \cite{2017Yajima} at redshift $z=7$. Whether the number density of such high mass objects are sufficiently high remains unknown.

\cite{2011Zackrisson} shows that 100 hr exposures at $z=10$ with \textit{JWST} should be able to detect Pop III galaxies of masses $M_\star \sim 5\times 10^5 \, \Msol$, having formed stars for 10 Myr, without the use of gravitational lensing. In our surveys we require similar masses for exposures of the same depth. Looking at the range $z=7-15$, \cite{2012Zackrisson} showed that \textit{JWST} imaging reaching $\sim 29.7$ in AB magnitude could find $\approx 1$ Pop III galaxy in one single NIRCam field of view when targeting well selected cluster lenses, in this case MACS J0717.5+3745, which provides higher probabilities for larger magnifications. These predictions were based on Pop III galaxies forming in atomic cooling halos with $\epsilon = 0.001$ and a top-heavy IMF with zero Lyman continuum leakage. As a comparison, using the randomly oriented wide or deep \textit{RST} surveys presented in this paper, we calculate that in order to find one Pop III galaxy at $z=10$ -- that is, only in one redshift bin $\Delta z=1$ -- \cite{2010Stiavelli} \& \cite{2020Visbal} predict number densities that are almost exactly at the requirements for photometric detection when using the star formation efficiency $\epsilon = 0.001$ -- the exact number of detections depends on whether the high or low mass data points are chosen in the simulations, but the difference is quite modest. Since we only consider one redshift bin at $z=10$, the number of detected Pop III galaxies would be increased by a factor of a few when integrated over the whole range of redshifts $z=7-15$ -- which would increase the number of detections by a few. Furthermore, large sky surveys conducted by e.g., \textit{RST} will most likely also include strong-lensing clusters -- but as shown, even without deliberately targeting cluster lenses, the wide and deep \textit{RST} photometric surveys require number densities that are more or less identical to the predictions by simulations at $z=10$, even for $\epsilon = 0.001$. As we discussed earlier, cluster lenses would offer a better alternative for spectroscopic detection of Pop III galaxies while photometry is arguably much more effective in a wide field survey like the ones planned with \textit{RST}.

\cite{2015Zackrisson} explored a scenario similar to the one in this paper by looking at the possibility of finding Pop III galaxies in randomly selected fields with magnifications up to $\mu \sim 1000$. They find that given an areal number density of $\approx 300$ objects arcmin$^{-2}$, a 100 deg$^2$ survey with a limiting magnitude of $m_\mathrm{AB} = 28$ would detect at least one Pop III galaxy of mass $\sim 1.5\times10^4\,\Msol$ at $z=7-13$ with magnifications $\mu \gtrsim 300$. In our predictions we find that the wide and deep \textit{RST} survey, covering $2000$ deg$^2$ at $m_\mathrm{AB} = 27$ and $40$ deg$^2$ at $m_\mathrm{AB} = 29$ respectively, have similar prospects for detection of a few Pop III galaxies without breaching the magnification limit ($\mu_\mathrm{max}=1000$) that we set for this paper. Once again, integration of the whole redshift range will increase the expected number of galaxies that will be detected by a factor of a few.

\subsection{The impact of variations in $\mathcal{L}_\mathrm{HeII \, 1640}$ and $\mathcal{L}_\mathrm{UV\,1500}$ } \label{ImpactOfLVar}
Other important factors to take into consideration are the expected deviations from the emission line fluxes and UV continuum predicted in, e.g., \cite{2010Raiter}. Combining eq.~\ref{eqPfit}, \ref{nmineq} and \ref{eqmuSpec} suggests that the minimum required number density has an inverse square dependence on the He II $\lambda$1640 $\ang$ emission line luminosity. This is valid for the range $\mu > 10$ in which the slope of the magnification function is $P(>\mu) \propto \mu^{-2}$. Therefore, a reasonable factor of $\approx 3$ variation in the luminosity of the He II $\lambda$1640 $\ang$ emission line will shift the required number density by a factor of $\approx 9$ for a given Pop III galaxy total stellar mass, therefore shifting the colored lines in the figures by an order of magnitude. Similarly, in the region of lower magnification $\mu \approx 1-10$, the magnification probability function has a steeper $\mu$ dependence and therefore introduces even larger variations which could be relevant for objects at lower redshift. Referring to figure~\ref{fig_Tele}a we see that an order of magnitude decrease in the minimum required number density cannot close that gap to the \cite{2010Stiavelli} simulations -- neither does it manage to bridge the gap to the simulations by \cite{2018Inayoshi,2020Visbal}. Looking at figure~\ref{fig_nminVSz} we see that across the redshift span presented, an order of magnitude decrease in the minimum required number density would render the \cite{2010Stiavelli} simulations detectable for $\epsilon = 0.1$ in the range $z\approx10-13$. In order for the fiducial \cite{2020Visbal} simulations to be made detectable (achieved by lowering the green, $\epsilon = 0.001$, line down to the green line with black square markers), one would require He II $\lambda$1640 $\ang$ emission line luminosities that are a factor of $\approx 100$ higher than predicted in the \cite{2010Raiter} stellar population synthesis models -- such variations should be regarded as highly unlikely.

Another important factor that would impact the He II $\lambda$1640 $\ang$ emission line luminosity is the escape fraction of He$^+$-ionizing photons. The escape fraction of ionizing radiation has proven to be very important for the strength of HI emission lines \citep[e.g.,][]{2014Wise}. However, the extent to which He$^+$-ionizing photons ($h\nu >54.4 $ eV) escape in the chemically pristine environments relevant for the birth Pop III galaxies, are not necessarily as severe as for metal enriched environments. \citet{2009Johnson} showed that only newly forming galaxies with quite high total stellar mass ($2.5\times 10^4 \ \Msol$) combined with a very top-heavy IMF (in that case a stellar population of 250 stars, all with mass $100 \ \Msol$) produces a significant escape fraction ($\sim 0.8$) of He$^+$-ionizing photons. In the case of lower total stellar mass ($2.5\times 10^3 \ \Msol$) and/or less top-heavy IMF, the escape fraction is entirely negligible -- which highlights the importance of the star formation efficiency and Pop III IMF. The simulations included in our paper spans a range of total stellar masses which is highly dependant on our choice of star formation efficiency. The \citet{2020Visbal} models have quite low star formation efficiency built into their models, and would therefore avoid the problem with a high escape fraction, while our fiducial choice for star formation efficiency in the \citet{2010Stiavelli} and \citet{2018Inayoshi} models produce Pop III galaxies with masses as high as $\sim 10^4 \ \Msol$ which could therefore stand the risk of suffering from a large He$^+$ ionizing escape fraction -- this would decrease the strength of the He II $\lambda$1640 $\ang$ emission line and subsequently require higher Pop III galaxy number densities. The top-heavy IMF used in our paper has a characteristic mass of $m_c = 60 \ \Msol$, $\sigma = 1$, and an allowed mass range of $1-500 \ \Msol$. For a total stellar mass of $2.5\times 10^4 \ \Msol$ this would produce $\sim 80$ stars with mass above $100\ \Msol$ -- which may result in a considerable He II $\lambda$1640 $\ang$ escape fraction according to \citet{2009Johnson}.

Similarly, variations in the continuum level at $1500\, \ang$ introduces the same order of magnitude changes in the minimum number density required for photometric detection -- as in the case for spectroscopic detection of the He II $\lambda$1640 $\ang$ emission line. Looking at figure~\ref{fig_Tele}b, we see that an order of magnitude downward shift of the colored lines would suggest that even the galaxies simulated in the fiducial \cite{2020Visbal} models could be detectable by wide/deep \textit{RST} photometric surveys -- whereas the number densities predicted by \cite{2018Inayoshi} continues to remain at a level much too low even for photometric detection.
\subsection{Simulation predictions vs. reality}
The accomplishment of this paper is unfortunately to have shown that we shouldn't expect to spectroscopically confirm Pop III galaxies in future planned surveys targeting large and randomly selected areas in the sky (random in the sense that the areas are not specifically targeted based on large magnifications from gravitational lensing). This however hinges on the predictions of contemporary simulations regarding the number density of atomic cooling halos and the expected total stellar mass of the Pop III galaxies embedded in them -- and our analysis helps to separate the detectable and undetectable parts of the parameter space. This hopefully compels the scientific community to further explore the possibilities of achieving high star formation efficiencies without making significant sacrifices with the number density of Pop III galaxies. At this moment we cannot say with certainty that such high efficiency models are ruled out and should therefore consider them seriously.

A candidate Pop III galaxy was recently reported by \cite{2020Vanzella}, showing strong Ly$\alpha$ emission at $z\approx6.6$, which may suggest that the lower star formation efficiencies provided by simulations are actually sufficient for spectroscopic detection of the Ly$\alpha$ emission lines. Applying the computational machinery used in this paper with the Ly$\alpha$ luminosity predicted in \cite{2010Raiter} for our fiducial lognormal top-heavy IMF ($m_c=60\,\Msol$, with mass range $1-500\,\Msol$), assuming 50 per cent transmission of Ly$\alpha$ through the intergalactic medium, at a source redshift of $z=7$ (the lowest redshift that we have included simulations for) -- our model shows that \cite{2020Visbal} predict number densities that are high enough to detect the Ly$\alpha$ emission line in $\sim 1$ Pop III galaxy of total stellar mass $\approx 4\times10^4 - 2\times10^5\,\Msol$ using the very deep \textit{JWST} survey. This neither defers nor confirms whether \cite{2020Vanzella} actually found a true Pop III galaxy. Rather, we see that finding lensed Ly$\alpha$ emitting Pop III galaxies close to the end of reionization should statistically be quite plausible, assuming that the transmission of Ly$\alpha$ through the intergalactic medium is sufficient.
\subsection{Other spectroscopic probes}
Having focused on the He II $\lambda 1640$ emission line as the best observational signature for detection of pristine Pop III galaxies, we have perhaps overlooked the utility of other emission lines that are useful in the search for Pop III galaxies. We mention the Ly$\alpha$ line as a very prominent emission line that could differentiate a Pop III galaxy from enriched stellar populations. This however restricts us to redshifts after reionization has completed due to the absorption of Ly$\alpha$ in the intergalactic medium. This does not render the line useless since we also note from figure~\ref{fig_nminVSz} that lower redshifts may indeed be very promising in the search for Pop III galaxies. As shown in \citep[e.g.,][]{2020LiuBromm}, small pockets of pristine gas is expected to remain after reionizations ends, enabling the study of Ly$\alpha$ emission from Pop III stars in some rare regions at lower redshift. Furthermore, starting with \textit{JWST} we will finally be able to probe the rest frame optical at the redshift frontier, such as the Balmer emission lines -- providing us with yet another useful tool in the search for Pop III. Being unaffected by the neutral intergalactic medium, the H$\alpha$ emission line will be within the wavelength range of \textit{JWST} up to $z\approx 7$ using NIRSpec, while the H$\beta$ emission line remains within range up to $z\approx 10$. These emission lines will generally be stronger than the He II $\lambda1640$ emission line, and will therefore be more easily detected at high redshift. For our fiducial IMF, star forming age ($\tau \approx 10$ Myr) at zero metallicity, we have that the ratio of hydrogen ionizing photons to He$^+$ ionizing photons are given by $\frac{Q_\mathrm{H}}{Q_\mathrm{He^+}} = 24.6$ for case B recombination \citep{2010Raiter,2002Schearer} -- suggesting that the H$\alpha$ and H$\beta$ luminosities are $\approx 5.3$ and 1.9 times higher than the He II $\lambda$1640 luminosity respectively -- however, these ratios will vary with IMF, age etc.. In the best circumstances (strongest HeII lines) we may find $\frac{Q_\mathrm{H}}{Q_\mathrm{He^+}} \sim 10$, leading to H$\alpha$ and H$\beta$ luminosities a factor of $\sim$ 2.2 stronger and a factor $\sim 0.77$ weaker respectively, when compared to the He II $\lambda$1640 luminosity. Furthermore, the He II $\lambda$4686 emission line can be used as a probe for a highly ionizing stellar population, but this line is a factor of $\sim 8$ weaker than the He II $\lambda$1640 emission line.

If very strong Balmer line equivalent widths are detected together with a complete lack of metal lines, one will have a good case for a Pop III galaxy or a galaxy with extremely metal poor chemical composition. \cite{2011Inoue} showed that extremely metal poor galaxies may be identified using constraints on the rest frame equivalent width (EW) of the Ly$\alpha$; (EW(Ly$\alpha$)>460 $\ang$), He II $\lambda$1640; (EW(He II)>1 $\ang$), O III; (EW(O III)<20 $\ang$), H$\alpha$; (EW(H$\alpha$)>3200 $\ang$) and H$\beta$; (EW(H$\beta$)>540 $\ang$) emission lines -- where the limits are stringent enough to avoid confusion with observations of very young ($\sim 1$ Myr) galaxies of higher metallicity. The O[III] $\lambda$5007 emission line would serve as the best indicator of metals when performing spectroscopy on selected targets. \citet{2011Inoue} continues to show that a ratio O[III]/H$\beta<0.1$ is sufficient to identify a galaxy as metal-free or extremely metal poor -- which can be confirmed for SFR $\sim 3 \ M_\odot$ galaxies up to $z\sim 8$ in deep exposures with NIRSpec on the \textit{JWST}.

\subsection{Spectroscopic follow-up}
Photometrically detected objects showing indicative Pop III color signatures may in principle be targeted for follow-up spectroscopy -- rendering the He II $\lambda$1640 emission line detectable for sufficiently deep exposures. As shown in \cite{2021Grisdale}, Pop III galaxies with total stellar masses of $\geq 10^4 \ M_\odot$, a top-heavy IMF and a compact stellar distribution would have a detectable He II $\lambda$1640 emission line for redshifts $z=4-10$ with the HARMONI spectrograph on the \textit{ELT}. In our paper we find that the sensitivity of the \textit{ELT} ($F_\mathrm{limit}\sim 10^{-19}$erg s$^{-1}$ cm$^{-2}$) is sufficient to allow spectroscopic follow-up on Pop III galaxies identified with deep \textit{RST} photometry across the whole relevant redshift range -- restricted only by the maximum magnification which limits us to $z\lesssim 11$ for the fiducial \citet{2020Visbal} while only placing a restriction on redshift for the high efficiency model and the \citet{2010Stiavelli} model due to redshifting out of the \textit{RST} filters (see figure~\ref{fig_Nvsz}).

\section{Conclusions} \label{conclusions}
Here, we summarize the key conclusions and take-home messages from this paper:
\begin{itemize}
\item Simulations predict number densities that are likely too low for spectroscopic detection of gravitationally lensed Pop III galaxies in wide-area surveys using \textit{JWST}, \textit{RST} and \textit{Euclid}. Allowing magnifications $\mu > 1000$, we find that all included telescopes provide similar prospects for spectroscopic detection. However, due to limitations in the highest applicable magnification when observing extended objects like galaxies, \textit{JWST} is the telescope that is best poised for spectroscopic detections of the He II $\lambda$1640 emission line. Photometric surveys will however, very likely, pick up lensed Pop III galaxies if star formation efficiencies are sufficiently high ($\epsilon \gtrsim 0.005$) but this is model dependent since not all simulations predict number densities that are sufficient for detection. When it comes to photometric detection of the UV continuum at 1500 $\ang$, the wide and deep \textit{RST} survey performs better than all other telescopes and surveys -- requiring Pop III galaxy number densities that are $\sim$ 2 orders of magnitude lower than required for detection in planned photometric \textit{JWST} surveys and $\sim 1.5$ orders of magnitude lower than required for the wide \textit{Euclid} survey (see e.g., figure~\ref{fig_Tele}). For the wide \textit{Euclid} survey we find minimum required number densities that are sufficient for photometric detections based on the \cite{2010Stiavelli} simulations for $\epsilon = 0.01$, although while generally breaching the maximum allowed magnification set in this paper for all included redshifts.

\item The star formation efficiency of Pop III star formation is paramount to the ability of detecting Pop III galaxies in gravitationally lensed fields due to its correlation with the total stellar mass formed, and therefore the brightness of Pop III galaxies. The poor prospects for spectroscopic detection of the He II $\lambda$1640 emission line, using our fiducial IMF and star forming age ($\tau$), can only be remedied if star formation efficiencies would reach $\epsilon \gtrsim 0.1$ while at the same time keeping the Pop III galaxy number density unchanged (see figure~\ref{fig_nminVSz}).
    
\item We conclude that the top-heavy lognormal IMF with characteristic mass $m_c = 60 \, \Msol$ and mass range $1-500 \, \Msol$ provide the best circumstances for spectroscopic detection among the IMFs included in this paper, but it still requires number densities about two orders of magnitude higher than the predictions from simulations. The less top-heavy IMFs included in this paper require number densities an additional $\sim 1.5$ orders of magnitude higher for spectroscopy (see figure~\ref{fig_IMFvar}) and $\sim 1$ orders of magnitude higher for photometry (section~\ref{ImpactofIMF}).

\item We find that during the first 10 Myr after the onset of star formation, the required number densities for spectroscopic and/or photometric detection varies very little, only at the level of a few per cent based on the luminosity of the galaxies. Looking instead at a shorter burst of star formation, e.g., 3 Myr, does however imply a brighter galaxy (higher SFR) given that the same total stellar mass is to be formed. This improves the detection prospects by lowering the minimum required number densities by about 1 order of magnitude while simultaneously lowering the number of pristine Pop III galaxies predicted by simulations by 0.5 orders of magnitude due to the shorter observational time window. This effectively reduces the gap between the simulations and our minimum required number density with $\sim 0.5$ orders of magnitude.

\item In this paper, simulations of Pop III galaxies come closest to the minimum required number densities for detection at $z\approx 10$ (see figure~\ref{fig_nminVSz} and section~\ref{theredshiftdepedence}). Whether our required minimum number density will coincide with the number densities predicted by simulations at lower redshift is not yet clear and therefore remains a topic of future research. 

\item Considering variations from the spectral synthesis model used in this paper \citep{2010Raiter}, we find that a factor of $\approx 3$ increase in the predicted He II $\lambda$1640 emission line luminosity corresponds to an order of magnitude decrease in the number density required for spectroscopic detection. At, e.g., $z=10$, this would not manage to render the \cite{2010Stiavelli} simulations detectable, assuming $\epsilon \approx 0.01$, and cannot bridge the gap to the simulations by \cite{2020Visbal} for reasonable variations in the emission line luminosity (see section~\ref{ImpactOfLVar}).

\item We make order of magnitude estimates which argues that cluster lensing surveys perform better than wide-field surveys for spectroscopic detection of Pop III galaxies. However, we argue that for photometric detection, the wide fields surveys perform considerably better. For every Pop III galaxy detected in a wide-field survey, we estimate that a single cluster lens would detect a factor of $\gtrsim 10$ more galaxies with spectroscopy. In the case of photometry one would instead require observations of several hundreds of cluster lenses for every galaxy detected in a wide-field survey in order to be competitive.

\item Gravitationally lensed Pop III galaxies detected in the photometric deep \textit{RST} survey would be sufficiently bright for spectroscopic follow-up on the He II $\lambda$1640 emission line using e.g., the \textit{ELT}.
\end{itemize}

\section*{Acknowledgements}
We thank the anonymous referee for comments that improved the scientific content of the paper. AV, EZ and MS acknowledge funding from the Swedish National Space Agency. EV acknowledges support from
NSF grant AST-2009309. MS acknowledges support from the Swedish Collegium for Advanced Study. EZ also acknowledges a sabbatical fellowship from AI4Research at Uppsala University.

\section*{Data availability}
The data underlying this article will be shared on reasonable request to the corresponding author.


\bibliographystyle{mnras}
\bibliography{bibliography} %



\bsp	
\label{lastpage}
\end{document}